\documentclass[iicol,pdflatex]{sn-jnl}% Style for submissions to Nature Portfolio journals
\usepackage{graphicx}%
\usepackage{multirow}%
\usepackage{amsmath,amssymb,amsfonts}%
\usepackage{amsthm}%
\usepackage{mathrsfs}%
\usepackage[title]{appendix}%
\usepackage{xcolor}%
\usepackage{textcomp}%
\usepackage{manyfoot}%
\usepackage{booktabs}%
\usepackage{algorithm}%
\usepackage{algorithmicx}%
\usepackage{algpseudocode}%
\usepackage{listings}
\usepackage{subfig}
\usepackage{afterpage}

\newcommand{\AlldsNumber}{64}
\newcommand{\alldataset}{64}

\newcommand{\SeqNumber}{336,476}
\newcommand{\pretrainMRI}{336,476}

\newcommand{\pretraindataset}{34}
\newcommand{\pretrainpublic}{8}
\newcommand{\pretrainprivate}{26}

\newcommand{\downstreamdataset}{37}
\newcommand{\downstreampublic}{32}
\newcommand{\downstreamprivate}{5}

\newcommand{\FTtasknumber}{44}
\newcommand{\downstreamtask}{44}

\newcommand{\PrivCenter}{15}

\newcommand{\modelname}{PRISM}

\newcommand{\modelFullName}{a foundation model \textbf{PR}e-trained with large-scale mult\textbf{I}-\textbf{S}equence \textbf{M}RI}

\theoremstyle{thmstyleone}%
\theoremstyle{thmstyletwo}%

\theoremstyle{thmstylethree}%

\begin{document}

\title[Article Title]{Large-scale Multi-sequence Pretraining for Generalizable MRI Analysis in Versatile Clinical Applications}

\author[1]{\fnm{Zelin} \sur{Qiu}}
\equalcont{These authors contributed equally to this work.}

\author[1]{\fnm{Xi} \sur{Wang}}
\equalcont{These authors contributed equally to this work.}

\author[2]{\fnm{Zhuoyao} \sur{Xie}}
\equalcont{These authors contributed equally to this work.}

\author[3,4]{\fnm{Juan} \sur{Zhou}}

\author[5]{\fnm{Yu} \sur{Wang}}

\author[5]{\fnm{Lingjie} \sur{Yang}}

\author[1]{\fnm{Xinrui} \sur{Jiang}}

\author[1]{\fnm{Juyoung} \sur{Bae}}

\author[1]{\fnm{Moo Hyun} \sur{Son}}

\author[2]{\fnm{Qiang} \sur{Ye}}

\author[2]{\fnm{Dexuan} \sur{Chen}}

\author[2]{\fnm{Rui} \sur{Zhang}}

\author[2]{\fnm{Tao} \sur{Li}}

\author[6]{\fnm{Neeraj Ramesh} \sur{Mahboobani}}

\author[7]{\fnm{Varut} \sur{Vardhanabhuti}}

% corresponding author

\author*[5]{\fnm{Xiaohui} \sur{Duan}} \email{duanxh5@mail.sysu.edu.cn}

\author*[2]{\fnm{Yinghua} \sur{Zhao}}\email{zhaoyh@smu.edu.cn}

\author*[1,8,9,10,11]{\fnm{Hao} \sur{Chen}}\email{jhc@cse.ust.hk}

\affil[1]{\orgdiv{Department of Computer Science and Engineering}, \orgname{The Hong Kong University of Science and Technology}, \orgaddress{\city{Hong Kong}, \country{China}}}

\affil[2]{\orgdiv{Department of Radiology}, \orgname{The Third Affiliated Hospital of Southern Medical University (Academy of Orthopedics, Guangdong Province)}, \orgaddress{\city{Guangzhou}, \country{China}}}

\affil[3]{\orgdiv{Department of Radiology}, \orgname{5th Medical Center of Chinese PLA General Hospital}, \orgaddress{\city{Beijing}, \country{China}}}

\affil[4]{\orgdiv{The Second School of Clinical Medicine}, \orgname{Southern Medical University}, \orgaddress{\city{Guangzhou}, \country{China}}}

\affil[5]{\orgdiv{Department of Radiology}, \orgname{Sun Yat-sen Memorial Hospital, Sun Yat-sen University}, \orgaddress{\city{Guangzhou}, \country{China}}}

\affil[6]{\orgdiv{Department of Imaging and Interventional Radiology, Faculty of Medicine}, \orgname{The Chinese University of Hong Kong}, \orgaddress{\city{Hong Kong}, \country{China}}}

\affil[7]{\orgdiv{Department of Diagnostic Radiology}, \orgname{Li Ka Shing Faculty of Medicine, The University of Hong Kong}, \orgaddress{\city{Hong Kong}, \country{China}}}

\affil[8]{\orgdiv{Department of Chemical and Biological Engineering}, \orgname{The Hong Kong University of Science and Technology}, \orgaddress{\city{Hong Kong}, \country{China}}}

\affil[9]{\orgdiv{Division of Life Science}, \orgname{The Hong Kong University of Science and Technology}, \orgaddress{\city{Hong Kong}, \country{China}}}

\affil[10]{\orgname{HKUST Shenzhen-Hong Kong Collaborative Innovation Research Institute}, \orgaddress{\city{Shenzhen}}, \country{China}}

\affil[11]{\orgdiv{State Key Laboratory of Nervous System Disorders}, \orgname{The Hong Kong University of Science and Technology}, \orgaddress{\city{Hong Kong}, \country{China}}}


\abstract{
Multi-sequence Magnetic Resonance Imaging (MRI) offers remarkable versatility, enabling the distinct visualization of different tissue types. This feature effectively outlines complex anatomical and pathological details, making it a fundamental diagnostic tool in a wide range of clinical situations. Nevertheless, the inherent heterogeneity among MRI sequences poses significant challenges to the generalization capability of deep learning models. These challenges undermine model performance when faced with varying acquisition parameters, thereby severely restricting their clinical utility. 
In this study, we present \modelname, \modelFullName. The model is designed to learn generalizable representations that adapt robustly to various clinical applications.
We collected a total of \alldataset~datasets from both public and private sources, encompassing a wide range of whole-body anatomical structures, with scans spanning diverse MRI sequences. Among them, \pretrainMRI~volumetric MRI scans from \pretraindataset~datasets (\pretrainpublic~public and \pretrainprivate~private) were curated to construct the largest multi-organ multi-sequence MRI pretraining corpus to date.
We propose a novel pretraining paradigm that disentangles anatomically invariant features from sequence-specific variations in MRI, while preserving high-level semantic representations. The framework combines pixel-level masked image reconstruction and image-to-image translation to maintain structural fidelity under varying contrast conditions. At the image level, metadata prediction is coupled with contrastive learning to enhance semantic representation learning. This dual-level disentanglement of anatomical priors from acquisition-dependent parameters reduces sensitivity to imaging protocols and significantly improves robustness to domain shifts across diverse MRI sequences.
For comprehensive validation, we established a benchmark comprising \downstreamtask~downstream tasks, including disease diagnosis, image segmentation, cross-sequence registration, progression prediction, and medical report generation.
These tasks were evaluated on \downstreampublic~public datasets and \downstreamprivate~private cohorts. \modelname~consistently outperformed both non-pretrained models and existing foundation models, achieving first-rank results in 39 out of 44 downstream benchmarks with statistical significance improvements.
These results underscore \modelname's ability to learn robust and generalizable representations across unseen data acquired under diverse MRI protocols. By bridging distributional discrepancies among heterogeneous MRI sequences, it maps contrast-specific representations into a unified semantic space. \modelname~provides a scalable framework for multi-sequence MRI analysis, thereby enhancing the translational potential of AI in radiology. It delivers consistent performance across diverse imaging protocols, reinforcing its clinical applicability.

}

\keywords{Representation Learning, Foundation Model, Large-scale Pretraining, Multi-sequence MRI}



\maketitle

\section{Introduction}\label{sec1}

Magnetic Resonance Imaging (MRI) is a foundational modality in modern clinical diagnostics, offering non-invasive, radiation-free visualization of soft tissue with excellent contrast resolution~\cite{katti2011magnetic, chavhan2008steady}. By modulating acquisition parameters, MRI enables the generation of diverse sequences, such as T1-weighted, T2-weighted, and proton density-weighted (PWD) images, each sensitized to different tissue properties~\cite{lundervold2019overview}. Clinical interpretation often requires synthesizing information across multiple sequences, but such multi-sequence MRI is high-dimensional, heterogeneous, and often incomplete, posing substantial interpretation challenges and demanding considerable expertise and time.

Deep learning has shown significant promise in automating medical image analysis~\cite{mazurowski2019deep, qiu2021predicting}. Yet, most existing approaches rely on supervised learning, which requires a large amount of labeled data, a major bottleneck in the clinical domain due to the high cost of expert annotation and privacy constraints. Moreover, these models are often tailored to specific organs or imaging protocols and generalize poorly across scanners, institutions, and patient populations~\cite{chen2021transunet}. To address annotation scarcity, transfer learning from natural images has been extensively explored. However, semantic and structural mismatches between natural and medical images severely limit the effectiveness of such approaches~\cite{raghu2019transfusion}. Furthermore, even pre-trained on medical datasets, such as Med3D~\cite{chen2019med3d}, many methods still require substantial fine-tuning with labelled data, limiting their scalability.

In recent years, foundation models tailored for radiological imaging have demonstrated strong performance across multiple modalities (e.g., X-ray, MRI, Computed Tomography) and a variety of anatomical sites, such as the lungs, liver, and heart~\cite{pai2024foundation,ying2024multicenter,wang2024screening}. However, extending these models to MRI-specific applications presents unique challenges.

First, while general-purpose medical foundation models have demonstrated a certain level of transferability to 3D MRI~\cite{ye2024continual}, fundamental differences in imaging physics, signal encoding, and distribution characteristics between MRI and other modalities can limit their effectiveness and hamper domain-specific generalization. 
%\cite{cox2024brainsegfounder, tak2024foundation, su2025slices}
Second, although tailored for MRI, existing domain-specific foundation models often suffer from limited anatomical coverage, inadequate exploitation of multi-sequence information, and narrow downstream task validation. For instance, BrainSegFounder~\cite{cox2024brainsegfounder} focused exclusively on brain imaging, restricted to T1-weighted and T2-weighted sequences, which constrains its applicability to more diverse clinical scenarios. Recent efforts have attempted to broaden anatomical and sequence representation. Triad~\cite{wang2025triad} was pre-trained on 3D MRI volumes from three anatomical regions and evaluated across multiple downstream tasks, like segmentation, classification, and registration. MRI-CORE~\cite{dong2025mri}, on the other hand, includes MRI scans from multiple anatomical regions for pretraining, but its reliance on 2D slice-based training limits the model’s ability to capture volumetric spatial context. In parallel, Sun et al.~\cite{sun2025foundation} proposed a foundation model targeting MR image enhancement via tissue-aware processing, demonstrating improvements in motion correction, denoising, and harmonization. Although downstream tasks are conducted on enhanced images, the model remains confined to enhancement objectives and lacks a task-agnostic representation suitable for broader clinical applications. Despite these advancements, several fundamental limitations remain across existing MRI foundation models.
First, anatomical diversity is still limited, as many models are trained on data from a single organ or restricted anatomical regions.
Second, multi-sequence information is often underutilized during pretraining, constraining the model’s ability to generalize across varying MRI sequences.
Third, robustness to real-world heterogeneity, such as variations in scanners, acquisition protocols, and patient populations, remains largely unexplored.

To address these challenges, we present \modelname, \modelFullName. With a novel pretraining paradigm, \modelname~is pre-trained on \pretrainMRI~multi-sequence MRI volumes (336k) collected from \pretrainpublic~public repositories and  \pretrainprivate~in-house database, covering 10 anatomical regions and various imaging protocols (Fig.~\ref{fig:pipeline}). \modelname~adopts a Swin Transformer~\cite{liu2021swin} backbone for hierarchical feature extraction and long-range dependency modeling, augmented by a novel dual-branch disentanglement module that explicitly separates anatomical features shared across sequences from sequence-specific contrast variations. To support generalization across scanners, protocols, and sequence availability, \modelname~is trained via a multi-task self-supervised learning framework that integrates four complementary objectives, including masked image reconstruction, cross-sequence translation, metadata prediction, and anatomy-invariant contrastive learning. These tasks jointly guide the model to learn robust and transferable representations that are anatomically consistent and invariant to MRI sequence variations.

%We systematically evaluate \modelname~on a large-scale benchmark, comprising \FTtasknumber~downstream tasks in a diverse range of clinical applications, including semantic segmentation, including pixel-level semantic segmentation (e.g., organ and tumor segmentation), image-level classification (e.g., abnormality detection, disease grading, sequence identification, and longitudinal progression forecasting), regression (e.g., age estimation), cross-sequence registration, and cross-modal radiology report generation. \modelname~consistently achieves state-of-the-art performance on all tasks on average, outperforming baselines with statistical significance (p value $<$ 0.001). Moreover, \modelname~has faster convergency, and stronger robustness against missing sequence issue. These results position \modelname~as a scalable, versatile, and clinically applicable foundation model for real-world multi-sequence MRI analysis.

We systematically evaluated \modelname~on a large-scale benchmark comprising \FTtasknumber~downstream tasks that span a wide range of clinical applications, including pixel-level semantic segmentation (e.g., organ and lesion segmentation), image-level classification (e.g., abnormality detection, disease grading, sequence identification, and longitudinal progression forecasting), regression tasks (e.g., age estimation), cross-sequence registration, and radiology report generation. \modelname~consistently achieved state-of-the-art performance in 39 of 44 downstream benchmarks, significantly outperforming strong baselines ($p<0.001$). In addition, \modelname~exhibits faster convergence and enhanced robustness to missing sequence scenarios. These results establish \modelname~as a scalable, versatile, and clinically applicable foundation model for real-world multi-sequence MRI analysis.

\begin{figure*}[htp!]
    \centering
    \includegraphics[width=\textwidth]{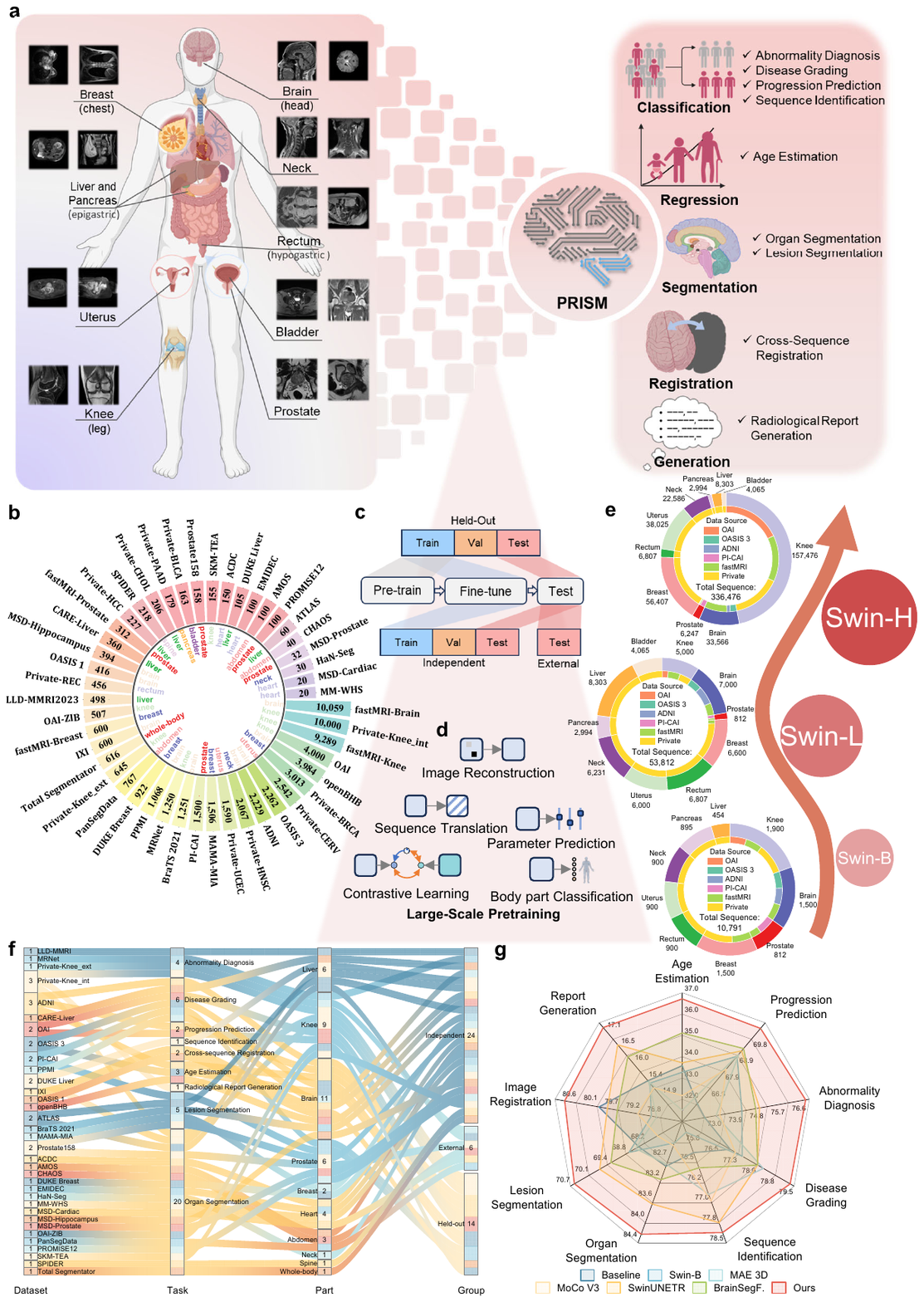}
\end{figure*}
\begin{figure*}[ht!]
    \centering
    \caption{
    Study overview.
    (a)  We propose \modelname, \modelFullName, which learns robust representations via pretraining on large-scale multi-sequence MRI data from diverse anatomical regions spanning from head to knee, thereby enabling strong generalization across nine radiological applications.  
    (b) The study utilizes \alldataset~datasets collected from both public and private sources. Each dataset is presented by its name, number of cases, and associated anatomical region. Notably, for visualization purposes, the private datasets used in pretraining have been merged according to their anatomical regions.
    (c) The downstream datasets are categorized into three evaluation groups: \textit{Held-out}, \textit{Independent}, and \textit{External} validation, reflecting increasing levels of distribution shift. 
    (d) Our pretraining framework integrates five pretext tasks: masked image reconstruction, sequence translation, acquisition parameter prediction, anatomical region classification, and contrastive learning.
    % During fine-tuning, pre-trained weights of Swin Transformer~\cite{liu2021swin}  are transferred to architectures including SwinUNETR~\cite{hatamizadeh2021swin}, MLP, TransMorph~\cite{chen2022transmorph}, and R2GenGPT~\cite{wang2023r2gengpt} for adapting to different tasks.
    (e) The scalability of \modelname~is evaluated by varying both the size of pretraining data (10k, 53k, 336k volumes) and model capacity (Swin-B, Swin-L, Swin-H). 
    (f) A systematic evaluation covering \FTtasknumber~tasks is conducted. This subfigure illustrates the relationships among each dataset, the anatomical regions they cover, the downstream tasks and assigned evaluation groups.
    (g) \modelname~is comprehensively evaluated using task-specific metrics: accuracy for classification, identification, grading, and progression prediction; Dice score for segmentation and registration; reciprocal of the mean absolute error (1/mean absolute error) for age prediction; and BLEU-1 score for report generation. The metrics are shown in percentages.
    }
    \label{fig:pipeline}
\end{figure*}
\section{Results}
In this study, we developed an MRI foundation model \modelname~for robust representation learning using a combined cohort of \pretrainMRI~multi-sequence MRI volumes (\modelname-336k) from \pretrainpublic~publicly available datasets and \pretrainprivate~in-house database, covering multiple anatomical regions, and various imaging protocols (Supplementary STable 4).
\modelname~was pre-trained through a novel self-supervised strategy, and then evaluated through fine-tuning by replacing the decoder with task-specific adapters tailored for versatile clinical applications, including segmentation, classification, regression, registration and generation, covering \downstreamtask~distinct downstream tasks.

To enable a systematic generalization analysis, the evaluation datasets were stratified into three cohorts, \textit{Held-out}, \textit{Independent}, and \textit{External}, based on their inclusion in the pretraining and fine-tuning stages (Fig.~\ref{fig:pipeline}(c)). The \textit{Held-out} cohort includes datasets whose training data were used in both pretraining and fine-tuning. The \textit{Independent} cohort comprises datasets whose training samples were used exclusively for fine-tuning, with no overlap with the pretraining data. The \textit{External} cohort contains datasets that were entirely excluded from both pretraining and fine-tuning, and used solely for evaluation. This configuration enables a rigorous evaluation of \modelname's generalization performance across domains exhibiting varying degrees of distributional overlap, from partially shared to entirely unseen data. In particular, for the five private knee datasets included in the downstream evaluation, Private-Knee\_K from center K was renamed Private-Knee\_int and utilized for fine-tuning, while Private-Knee-L, Private-Knee-M, Private-Knee-N, and Private-Knee-O were merged into a single dataset within the \textit{External} cohort (named Private-Knee\_ext) for evaluation.

For comparison, we included a diverse set of state-of-the-art (SOTA) models spanning three categories: two task-specific supervised baselines, i.e., ResNet 3D~\cite{he2016deep}, nnUNet~\cite{isensee2021nnu}; two general-purpose self-supervised learning (SSL) methods, i.e., MAE~\cite{he2022masked}, MoCo V3~\cite{chen2021empirical}; and two MRI foundation models, i.e., SwinUNETR~\cite{tang2022self}, BrainSegFounder~\cite{cox2024brainsegfounder} (BrainSegF. for short in this work). 
To assess scalability and investigate scaling behavior, we derived two subsets from the full 336k dataset (10k and 53k samples; Fig.~\ref{fig:pipeline}(e)) and conducted fine-tuning across multiple model sizes (Swin-B, Swin-L, and Swin-H) under varying data regimes. An overview of the study is shown in Fig.~\ref{fig:pipeline}.

\subsection{Semantic Segmentation}
\begin{figure*}
    \centering
    \includegraphics[width = \textwidth]{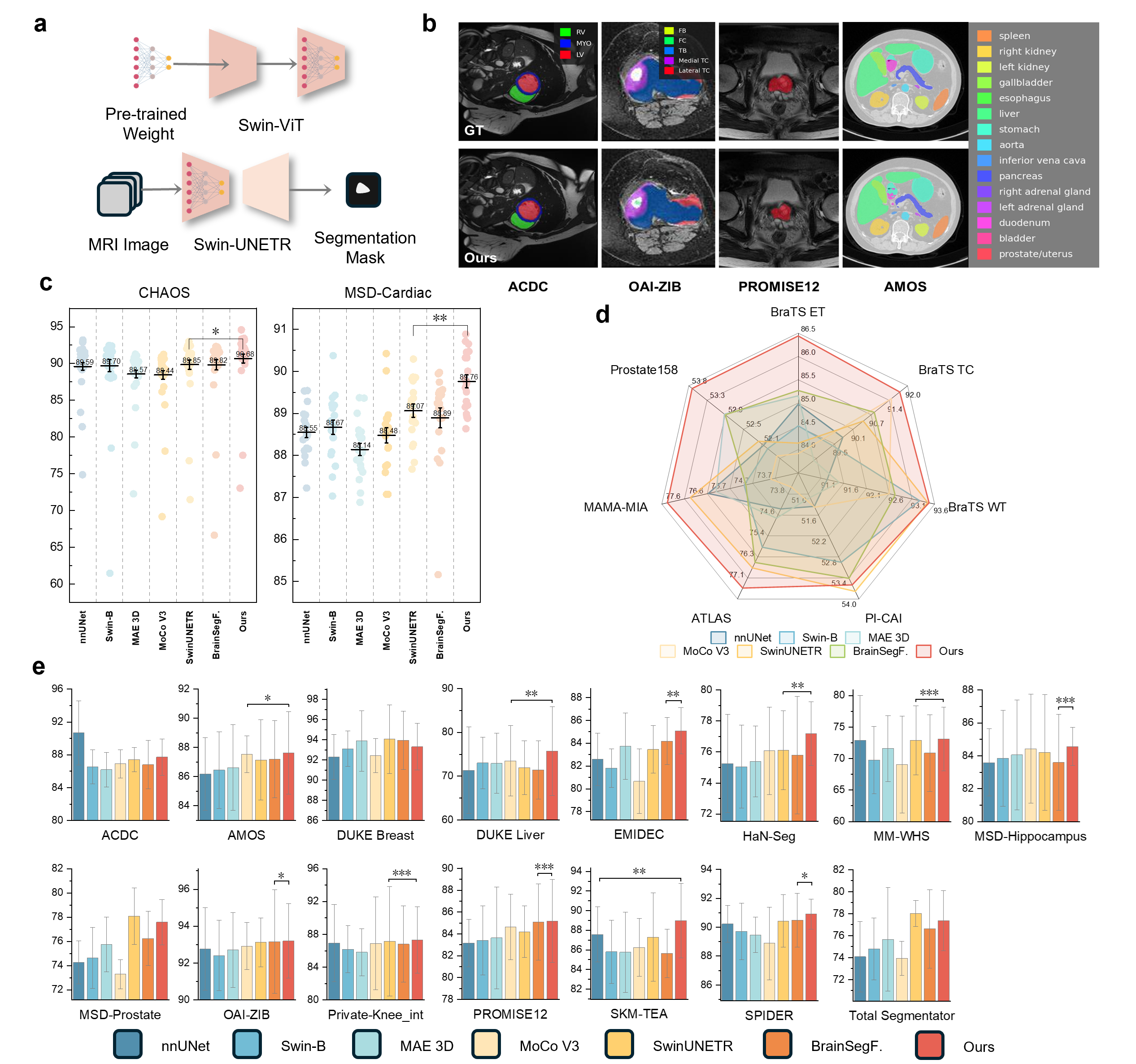}
    \caption{Evaluation on semantic segmentation. 
    (a) illustrates the adaptation of pre-trained Swin Transformer (Swin-ViT) weights to the encoder of the SwinUNETR architecture, which takes MRI images as input and generates segmentation masks.
    (b) shows the qualitative results of the \modelname~ on organ segmentation model (bottom panel) in comparison with the ground truth annotations (upper panel) across four representative tasks: heart structure segmentation on the ACDC dataset, knee structure segmentation on OAI-ZIB, prostate segmentation on the PROMISE12 dataset, and multi-organ abdominal segmentation on the AMOS dataset.
    %Demonstrates the effectiveness of our downstream organ segmentation approach. Our model can handle the segmentation of multiple organs in complex imaging sequences while maintaining high performance.  
    (c) \textit{External} validation on the abdomen organs (CHAOS) and heart (MSD-Cardiac) datasets shows that our method is more generalizable to unseen data than other methods. We report Dice score in segmentation tasks. The significance of differences between our method and the second-best method is marked with *, **, and ***, indicating $p<0.05$, $p<0.01$, and $p<0.001$, respectively. 
    (d) Performance comparison in lesion segmentation tasks, including brain tumor (BraTS), breast cancer (MAMA-MIA), and prostate cancer (PI-CAI). The surrounding lines highlight that our model consistently outperforms the comparison models.
    (e) The Dice results across organ segmentation tasks. Error bars represent the 95\% confidence interval (CI). }
    \label{fig:segmentation}
\end{figure*}

Semantic segmentation is a fundamental task in medical image analysis, enabling voxel-wise delineation of anatomical and pathological structures in MRI scans. We evaluate \modelname~on two major categories of segmentation tasks: organ segmentation and lesion segmentation. These tasks span multiple anatomical sites and MRI sequences, evaluating the model’s ability to learn structure-consistent, contrast-invariant representations from 3D volumetric data. Fig.~\ref{fig:segmentation}(a) illustrates the pipeline for adapting our model to segmentation tasks. Specifically, we initialize the encoder of SwinUNETR with pre-trained weights from the Swin Transformer (Swin-ViT). This resulting model takes MRI images as input and generates corresponding segmentation masks.

\subsubsection{Organ Segmentation}
We evaluated \modelname~on 20 MRI datasets for organ segmentation, including 15 datasets with finetuning and 5 for \textit{External} validation. The datasets encompassing a broad anatomical spectrum including the heart (ACDC~\cite{dsACDCbernard2018deep}, EMIDEC~\cite{dsEMIDEClalande2020emidec}, MSD-Cardiac~\cite{dsMSDantonelli2022medical}, MM-WHS~\cite{dsMMWHSzhuang2018multivariate}), hippocampus (MSD-hippocampus~\cite{dsMSDantonelli2022medical}), breast (DUKE Breast~\cite{dsDUKEBreastsaha2018machine}), abdominal organs (AMOS~\cite{dsAMOSji2022amos}, ATLAS~\cite{dsATLASquinton2023tumour}, CHAOS~\cite{dsCHAOS2021}, DUKE Liver~\cite{dsDUKELivermacdonald2023duke}, PanSegData~\cite{dspansegzhang2025large}), prostate (MSD-Prostate~\cite{dsMSDantonelli2022medical}, PROMISE12~\cite{dsPROMISE12litjens2014evaluation}, Prostate158~\cite{dsProstate158adams2022prostate158}), knee (Private-Knee\_int, OAI-ZIB~\cite{dsOAIZIBambellan2019automated}, SKM-TEA~\cite{dsSKMTEAdesai2022skm}), spine (SPIDER~\cite{dsSPIDERvan2024lumbar}), neck (HaN-Seg~\cite{dsHaNpodobnik2023han}), and whole-body organs (Total Segmentator~\cite{dsTotalSegd2024totalsegmentator}). Among these, Private-Knee\_int and OAI-ZIB were assigned to the \textit{Held-out} cohorts, while ATLAS, MSD-Cardiac, PanSegData, Private-Knee\_ext, and Prostate158 were assigned to the \textit{External} cohorts. The remaining datasets were included in the \textit{Independent} cohorts.

As shown in Fig.~\ref{fig:segmentation}(e) and Supplementary STable 5, \modelname~achieved the highest Dice scores on both \textit{Held-out} cohorts, reaching 87.30\% on Private-Knee\_int and 93.20\% on OAI-ZIB. Across the remaining \textit{Independent} cohorts, our method achieved the highest Dice scores on 9 out of 13 datasets, including AMOS (87.63\%), MSD-Hippocampus (84.57\%), PROMISE12 (85.18\%), SPIDER (90.91\%), HaN-Seg (77.18\%), EMIDEC (85.09\%), DUKE Liver (75.70\%), MM-WHS (73.12\%), and SKM-TEA (88.99\%). On these tasks, it outperformed the second-best method by up to 2.20\% (DUKE Liver), with consistently strong results across both SSL and foundation models. Notably, \modelname~surpassed the Swin-B baseline across all 13 tasks significantly (p$<$0.001), with Dice improvements ranging from 0.21\% to 3.32\%.%, and an average gain of \textbf{1.96\%}.

%To further assess out-of-distribution generalization, we evaluated \modelname~on five \textit{External} datasets: CHAOS, MSD-Cardiac, Prostate158, ATLAS, and PanSegData. We adopted a zero-shot transfer setting, where models fine-tuned on semantically related source datasets were directly applied to the external datasets without any additional adaptation. Specifically, the AMOS model was transferred to CHAOS, ATLAS, and PanSegData; the DUKE Breast model to MSD-Cardiac; and the MSD-Prostate model to Prostate158. As shown in Fig.~\ref{fig:segmentation}(c), \modelname~achieved the highest Dice scores across two external cohorts with small data and maintaining low division. Compared to Swin-B, our method yielded absolute improvements of 1.09\% to 2.52\%, with an average gain of 1.68\%. 
%Compared to MAE and MoCo V3, \modelname~demonstrated consistent advantages in zero-shot generalization as well—for example, on CHAOS, it improved over MAE and MoCo V3 by 1.86\% and 2.24\%, respectively; on PanSegData, the improvements were 2.20\% and 1.46\%. In addition, it outperformed the second-best methods on all external datasets, achieving an average Dice improvement of 0.49\%.
To further assess out-of-distribution generalisation, we evaluated \modelname~on five \textit{External} datasets: CHAOS, MSD-Cardiac, Prostate158, ATLAS, and PanSegData. We adopted a zero-shot transfer setting, where models fine-tuned on semantically related source datasets were directly applied to the \textit{External} datasets without any additional adaptation. Specifically, the AMOS model was transferred to CHAOS, ATLAS, and PanSegData; the MM-WHS model to MSD-Cardiac; and the PROMISE12 model to Prostate158. As illustrated in Fig.~\ref{fig:segmentation}(c) and the Supplementary STable 9, \modelname~consistently achieved the highest Dice scores across these \textit{External} datasets. Compared to Swin-B, our method yielded absolute improvements of 1.09\% to 2.52\%, with an average gain of 1.68\%. In comparison to MAE and MoCo V3, \modelname~also exhibited robust zero-shot generalization. For example, on CHAOS, it outperformed MAE and MoCo V3 by 1.86\% and 2.24\%, respectively; on PanSegData, the corresponding improvements were 2.20\% and 1.46\%. Furthermore, across all \textit{External} datasets, \modelname~consistently surpassed the second-best method. 
Although the absolute improvements were sometimes modest, this might be attributed to their relatively limited anatomical complexity, small domain gap, or well-curated annotations in certain datasets. In such scenarios, multiple models, including strong SSL and supervised baselines, can already achieve competitive results, leaving limited room for further improvement. Nevertheless, the consistent top-ranking performance across all \textit{External} benchmarks confirms the strong out-of-distribution generalization of \modelname, even when the task difficulty or domain gap is relatively low.

Importantly, although \modelname~slightly underperformed nnUNet or SwinUNETR on ACDC and MSD-Prostate, respectively, in the \textit{Independent} evaluation setting, our model achieved better results on the \textit{External} datasets with the same anatomical region, improving Dice by 1.09\% on MSD-Cardiac and 1.44\% on Prostate158. These findings suggest that \modelname~offers greater robustness to real-world data variability, with strong generalization ability and high potential for zero-shot deployment across diverse clinical scenarios without task-specific fine-tuning. More granular segmentation results, reporting Dice scores for individual heart, knee, and spine structures, are provided in Supplementary STables 7–9.

% \begin{figure*}
%     \centering
%     \includegraphics[width = \textwidth]{figures/organSeg.png}
%     \caption{We report Dise score (DSC) in organ segmentation tasks. We compare the model with Swin-B which trained from scratch and label the improvement on the top of each column. We marked the significance of differences between our method and the second-best method, with *,**, and *** indicates $p<0.05$, $p<0.01$ and $p<0.001$, respectively.}
%     \label{fig:org_seg}
% \end{figure*}

% \begin{figure}
%     \centering
%     \includegraphics[width=0.5\textwidth]{figures/sag_ext.png}
%     \caption{Comparison of segmentation performance on external datasets.}
%     \label{fig:seg_ext}
% \end{figure}

% \begin{figure*}[ht]
%     \centering
%     \includegraphics[width=0.9\textwidth]{figures/seg_vis.png}
%     \caption{Qualitative results on organ segmentation. }
%     \label{fig:seg_vis}
% \end{figure*}

\subsubsection{Lesion Segmentation}

We further evaluated \modelname~on five lesion segmentation tasks, including brain tumor (BraTS~\cite{dsBraTSbaid2021rsna}), breast cancer (MAMA-MIA~\cite{dsmamamiagarrucho2025}), prostate cancer (PI-CAI~\cite{dsPICAIsaha2024artificial} and Prostate158~\cite{dsProstate158adams2022prostate158}), and liver metastasis (ATLAS~\cite{dsATLASquinton2023tumour}). Among these, PI-CAI was used as a \textit{Held-out} cohort, while the remaining datasets served as \textit{Independent} cohorts.
As shown in Fig.~\ref{fig:segmentation}(d) and Supplementary STable 10, \modelname~achieved the highest Dice scores across all five tasks, reaching 90.58\% on BraTS, 77.56\% on MAMA-MIA, 77.14\% on PI-CAI, 88.76\% on Prostate158, and 77.38\% on ATLAS.
Compared to the supervised baseline Swin-B, \modelname~achieved consistent improvements, ranging from 1.25\% on PI-CAI to 3.26\% on MAMA-MIA, with an average gain of 1.99\% across all tasks. The large improvement on MAMA-MIA suggests enhanced discriminative capability in segmenting small or heterogeneous lesions under domain shift.
Moreover, while SwinUNETR and BrainSegF. showed competitive performance, \modelname~surpassed both, by margins of 0.75\% on BraTS and 0.78\% on ATLAS, two particularly challenging tasks due to substantial inter-subject variability and heterogeneous imaging quality.
More granular segmentation results, reporting Dice scores for tumor sub-regions, are provided in Supplementary STables 11.

% TODO: change, add ext prostate158
% \begin{figure}
%     \centering
%     \includegraphics[width=0.5\textwidth]{figures/segradar.png}
%     \caption{Performance comparison of cancer-related segmentation. Different lines and colors corresponding to each model, our method has surpasses other methods in all the sub tasks.}
%     \label{fig:seg_int}
% \end{figure}

\subsection{Image Classification}
To comprehensively assess the classification performance of our model, we conducted experiments on four clinically relevant tasks. We began with MRI sequence classification to evaluate the model’s ability to distinguish different MRI sequence types. We then assessed abnormality diagnosis through multiple binary classification tasks targeting specific pathological findings. Next, we evaluated disease grading as a multi-class task to quantify disease severity. Finally, we tested the model on a disease progression forecasting task to predict longitudinal changes in clinical status. 

\begin{figure*}
    \centering
    \includegraphics[width=0.9\textwidth]{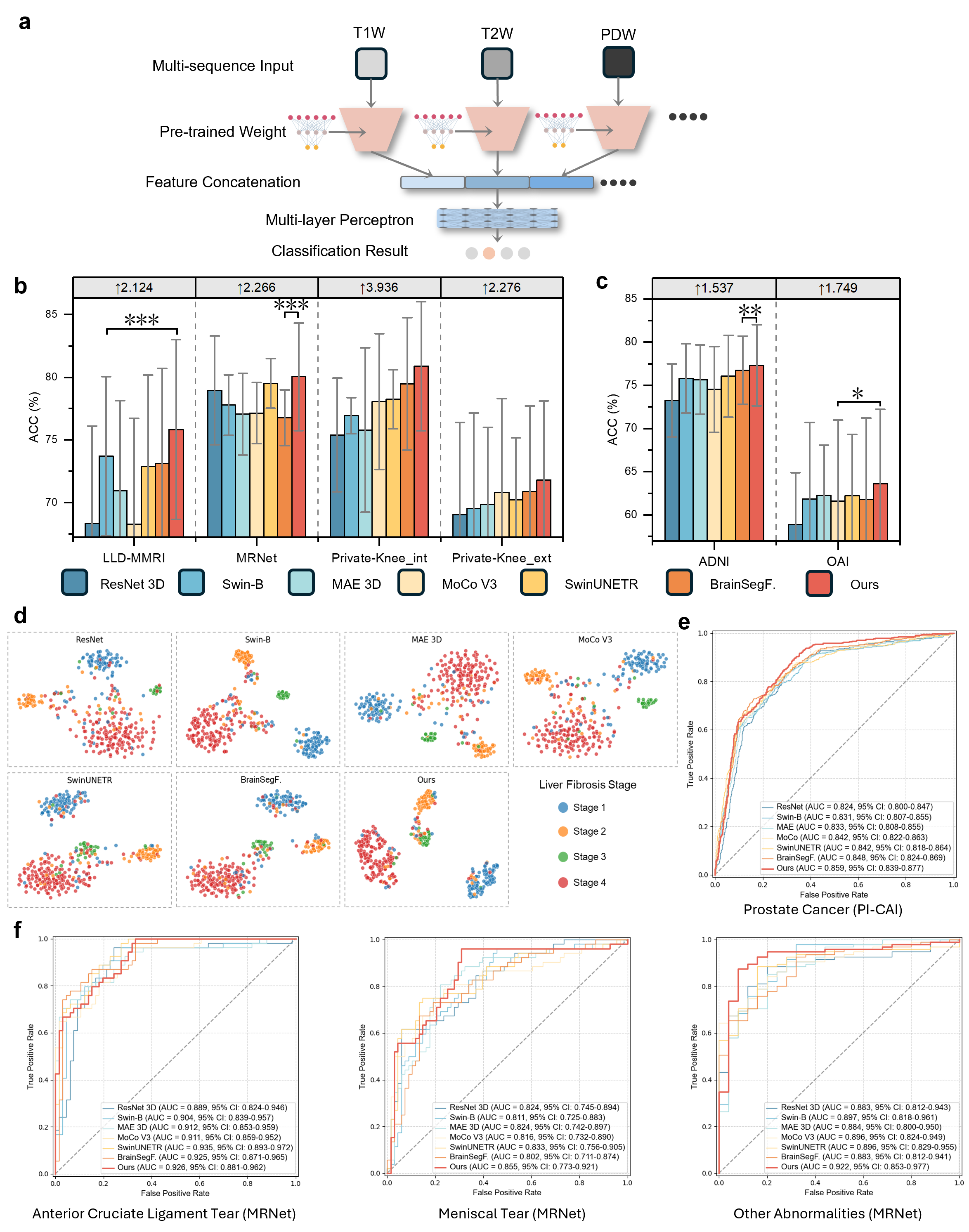}
    \caption{Evaluation on different classification tasks. 
    (a) We adapt multiple Swin-ViT encoders for multi-sequence input, with each encoder dedicated to process a specific sequence. The extracted features are then concatenated and fed into a multi-layer perceptron (MLP) to generate the final classification output.
    (b) The accuracy of abnormality diagnosis is evaluated on LLD-MMRI~\cite{dsLLDlou2025sdr}, MRNet~\cite{dsMRNetbien2018deep}, Private-Knee\_int, and Private-Knee\_ext, which correspond to two \textit{Independent}, one \textit{Held-out}, and one \textit{External} validation sets. The numbers on top denote the performance improvement compared to the unpretrained Swin-B model. The statistical significance between the best model and the second-best model is marked as: *(p$<$0.05), ** (p$<$0.01) and *** (p$<$0.001), respectively. Notably, pretraining leads to reduced variance and enhanced accuracy in out-of-distribution testing on the \textit{External} dataset, highlighting its generalization capability.
    (c) For disease progression prediction, the model determines the upcoming stages of Alzheimer’s disease (ADNI) and osteoarthritis (OAI). The error bars represent the 95\% confidence interval (CI), illustrating the stability of the prediction results across multiple trials.
    (d)We visualize the features from different liver fibrosis stages on the CARE-Liver dataset, with each colour representing one stage. The features extracted by different models are reduced dimensionality to 2D using t-SNE~\cite{maaten2008visualizing}. Compared to baselines, our model yields more compact and discriminative clusters, demonstrating its superior capability to capture fine-grained distinctions in pathological features.
    (e)(f) For the ROC curves of prostate cancer (PI-CAI) and knee abnormalities (MRNet), the area under the receiver operating characteristic curve (AUC-ROC) metric with 95\% confidence intervals (CI) is reported in the lower right legend.
    }
    \label{fig:classification}
\end{figure*}

\subsubsection{Abnormality Diagnosis}

We evaluate abnormality diagnosis performance on a multi-phasic liver MRI dataset (LLD-MMRI) and three knee MRI datasets. LLD-MMRI, categorized as an \textit{Independent} cohort, comprises seven diagnostic labels and is formulated as a multi-label classification task to capture co-occurring liver abnormalities. The knee datasets span different evaluation settings, including an \textit{Independent} cohort (MRNet~\cite{dsMRNetbien2018deep}) with three diagnostic labels, a \textit{Held-out} cohort (Private-Knee\_int), and an \textit{External} cohort (Private-Knee\_ext), each containing twelve binary labels under a multi-label classification framework. Further dataset details are provided in the Supplementary file.

As shown in Fig.~\ref{fig:classification}(b) and Supplementary STable 12, \modelname~achieved the highest classification accuracy across all knee MRI datasets, with 80.05\% on MRNet, 80.88\% on Private-Knee\_int, and 71.82\% on Private-Knee\_ext. Compared to the task-specific supervised baseline (Swin-B), it yielded absolute improvements of 2.27\%, 3.94\%, and 2.28\% on the \textit{Independent}, \textit{Held-out}, and \textit{External} cohorts, respectively.
While MAE and MoCo V3 performed competitively on the public MRNet dataset, their performance degraded substantially on private datasets under domain shift. For instance, MAE dropped to 69.85\% on Private-Knee\_ext. In contrast, \modelname~consistently outperformed both self-supervised methods and existing foundation models (e.g., SwinUNETR, BrainSegF.), demonstrating superior robustness and generalization in both in-domain and out-of-distribution settings. Furthermore, it significantly outperformed baseline methods across tasks (all $p<0.01$), indicating consistent statistical superiority.

% \begin{figure}
%     \centering
%     \includegraphics[width=0.5\textwidth]{figures/abn_cls.png}
%     \caption{}
%     \label{fig:abn_cls}
% \end{figure}

\subsubsection{Disease Grading}
We further evaluated \modelname~on six disease grading tasks, across a diverse range of clinical cohorts, including two \textit{Independent} datasets, i.e., PPMI (Parkinson’s disease) and CARE-Liver (liver fibrosis), and four \textit{Held-out} cohorts, i.e., OAI (knee osteoarthritis), ADNI and OASIS 3 (Alzheimer’s disease), and PI-CAI (prostate cancer). As shown in Table~\ref{tab:dis_grading} and Supplementary STable 13, \modelname~achieved the highest accuracy on five out of six datasets, reaching an accuracy of 71.75\% on PPMI, 
79.86\% on CARE-Liver, 64.20\% on OAI, 85.05\% on OASIS 3, and 86.78\% on PI-CAI. Compared to the supervised baseline Swin-B, \modelname~consistently outperformed it across all tasks, with statistically significant improvements ranging from 1.00\% (OAI, $p<0.01$) to 2.63\% (PPMI, $p<0.05$), and an average gain of 1.93\%. 
Notably, the osteoarthritis grading in OAI follows previous work~\cite{guida2021knee}, where assigning Kellgren-Lawrence (KL) grades posed considerable difficulty due to the subtle nature of imaging cues when relying solely on pure MRI images. This highlights the model's ability to extract discriminative features from complex medical imaging data, even in scenarios with ambiguous or subtle diagnostic markers.

While MAE and MoCo V3 performed competitively on some tasks (e.g., MAE reached 88.41\% on ADNI), their performance dropped substantially on others. For instance, MAE only achieved 69.97\% on PPMI compared to \modelname's 71.75\%, respectively, suggesting limited robustness across cohorts with varying anatomical or phenotypic characteristics. Although BrainSegF. slightly outperformed \modelname~on ADNI (0.61\%), our method achieved statistically significant gains over the second-best models on the other five tasks, with improvements ranging from 0.03\% to 1.68\% (mean: 0.57\%), all of which were statistically significant ($p<0.05$). These findings underscore the strong discriminative capability and robustness of \modelname~in diverse grading scenarios, particularly in settings characterized by subtle phenotypic differences or domain shifts.
Fig.~\ref{fig:classification}(d) visualizes the feature distributions obtained by different methods on the CARE-Liver dataset~\cite{dscare3wu2022meru} using t-SNE dimensionality reduction~\cite{maaten2008visualizing}, with colors representing different fibrosis stages. \modelname~exhibits noticeably tighter and more compact clusters, particularly for Stages 1 and 4, with fewer instances overlapping across classes. This suggests that our model captures more fine-grained and discriminative features while effectively minimizing intra-class variation.

\begin{table*}[]
    \centering
    \caption{The accuracy (\%) in disease grading tasks for Parkinson's disease (PPMI), knee osteoarthritis (OAI), Alzheimer’s disease (ADNI and OASIS 3) and prostate cancer (PI-AI). ResNet3D and Swin-B are trained from scratch. The last row presents the relative gains of \modelname~compared to Swin-B trained from scratch, with stars in brackets denoting statistical significance (* for $p<0.05$, ** for $p<0.01$, and *** for $p<0.001$).}
    \resizebox{\textwidth}{!}{
    \begin{tabular}{lllllll}
    \toprule
    
    Model\textbackslash   Dataset & ADNI~\cite{dsADNImueller2005alzheimer} & CARE-Liver~\cite{dscare3wu2022meru}& OAI~\cite{dsOAInevitt2006osteoarthritis} & OASIS 3~\cite{dsOASISlamontagne2019oasis} & PI-CAI~\cite{dsPICAIsaha2024artificial} & PPMI~\cite{dsPPMImarek2018parkinson} \\
    \midrule
    ResNet3D                      & 87.73 (±3.86)             & 74.20 (±4.27)             & 62.04 (±2.20)             & 83.04 (±3.87)             & 85.23 (±4.86)             & 69.38 (±4.44)             \\
    Swin-B                        & 86.37 (±3.56)             & 78.11 (±4.92)             & 63.20 (±4.48)             & 82.74 (±3.36)             & 84.94 (±6.37)             & 69.13 (±4.27)             \\
    MAE 3D                        & 88.41 (±2.49)             & 78.01 (±4.91)             & \underline{64.08 (±3.01)} & \underline{83.98 (±2.27)} & 85.14 (±3.98)             & 69.97 (±3.83)             \\
    MoCo V3                       & 86.04 (±2.40)             & 76.54 (±4.86)             & 63.18 (±2.59)             & 81.31 (±3.99)             & 85.81 (±5.43)             & 69.95 (±3.37)             \\
    SwinUNETR                     & 87.40 (±1.78)             & \underline{78.18 (±3.62)} & 62.05 (±3.58)             & 82.82 (±3.46)             & \underline{86.75 (±6.01)} & \underline{71.00 (±3.22)} \\
    BrainSegF.                    & \textbf{89.03 (±1.87)}    & 77.24 (±5.73)             & 62.29 (±4.03)             & 83.54 (±3.38)             & 86.21 (±7.16)             & 69.90 (±2.23)             \\
    Ours                          & \underline{88.42 (±2.17)} & \textbf{79.86 (±4.38)}    & \textbf{64.20 (±2.90)}    & \textbf{85.05 (±2.48)}    & \textbf{86.78 (±7.39)}    & \textbf{71.75 (±2.11)}    \\
    \midrule
    $\Delta$ (Swin-B)             & 2.05 (*)                  & 1.76 (***)                & 1.00 (**)                 & 2.31 (***)                & 1.84 (**)                 & 2.63 (*)                  \\ 
    \bottomrule
    \end{tabular}
    }
    \label{tab:dis_grading}
\end{table*}

\subsubsection{Progression Prediction}

We assessed the efficacy of \modelname~in modeling disease progression using two widely studied longitudinal cohorts—ADNI for Alzheimer’s disease and OAI for osteoarthritis.
Leveraging baseline images and subsequent scans at 12- or 24-month intervals, the model accurately classifies progression stages for upcoming timepoints.
As summarized in Fig.~\ref{fig:classification}(c) and Supplementary STable 14, \modelname~achieved the highest accuracy on both tasks, with 77.34\% on ADNI and 63.62\% on OAI. Compared to the Swin-B baseline, our method achieved statistically significant improvements of 1.54\% ($p<0.001$) and 1.75\% ($p=0.064$), respectively. While other foundation models such as BrainSegF. and SwinUNETR also demonstrated competitive performance, \modelname~consistently outperformed all compared models significantly. 
The result demonstrates its capability to extract and integrate key features of subtle changes across different sequences, enabling effective prediction of disease trajectories in distinct pathological contexts.

\subsubsection{MRI Sequence Identification}
We further evaluated \modelname~on the sequence identification task using the DUKE Liver dataset. As shown in Supplementary STable 15, \modelname~achieved the highest classification accuracy of 78.21\% (95\% confidence intervals [CI]: 73.05-83.38), outperforming all baseline models. The second-best performance was obtained by SwinUNETR at 77.84\% (95\% CI: 73.09-82.59), while the supervised baseline Swin-B yielded 74.41\% (95\% CI: 69.52-79.30). Compared to Swin-B, \modelname~demonstrated an absolute improvement of 3.80\%, which was statistically significant ($p<0.001$). Other self-supervised methods, including MAE (76.76\%), MoCo V3 (77.01\%), and BrainSegF. (75.75\%), showed lower accuracies and wider confidence intervals. These results underscore the superior capability of \modelname~to capture sequence-specific representations, and its enhanced generalization in discriminating subtle differences across MR acquisition protocols.

\subsection{Age Estimation}
Regression tasks offer greater flexibility for modeling continuous phenotypic variation compared to classification or grading tasks. To evaluate the generalizability of \modelname~in such settings, we conducted age prediction on three brain MRI datasets: ADNI~\cite{dsADNImueller2005alzheimer}, OASIS 3~\cite{dsOASISlamontagne2019oasis}, and OpenBHB~\cite{dsOpenBHBdufumier2022openbhb}. As shown in Table~\ref{tab:age_pred}, \modelname~consistently achieved the lowest mean absolute error across all cohorts, i.e., 2.48 on ADNI, 2.28 on OASIS 3, and 3.42 on OpenBHB, outperforming all baselines.

Compared to the supervised Swin-B baseline, \modelname~significantly reduced prediction error by 0.31 on ADNI ($p<0.01$), 0.31 on OASIS3 ($p<0.05$), and 0.23 on OpenBHB ($p<0.001$). Other pre-trained models such as BrainSegF. and SwinUNETR achieved moderate gains over from-scratch baselines (e.g., BrainSegF.: 0.260 on ADNI, 0.009 on OASIS3, and 0.203 on OpenBHB), but consistently underperformed relative to \modelname. On the most heterogeneous dataset, OpenBHB, which features an age-balanced design including younger individuals, \modelname~reduced the mean absolute error by 0.026 compared to BrainSegF. (3.443) and by 0.27 compared to SwinUNETR (3.69), highlighting its superior robustness for continuous phenotypic modeling in neuroimaging applications.

\begin{table*}[ht]
    \centering
    \caption{We report the mean absolute error (95\% CI) in the age estimation task, where a lower value indicates better performance. The best-performing model is denoted in bold font, and the second-best model is underlined. The last row presents the relative gains of \modelname~compared to Swin-B trained from scratch, with stars in brackets denoting statistical significance (* for $p<0.05$, ** for $p<0.01$, and *** for $p<0.001$).}
    \label{tab:age_pred}
    \resizebox{\textwidth}{!}{
    \begin{tabular}{llll}
    \toprule
    Model\textbackslash   Dataset & ADNI                                       & OASIS 3                                     & OpenBHB                                   \\ \midrule
    ResNet3D                      & 2.652 (1.702-3.602), p$<$0.05              & 2.519 (1.486-3.553), p$<$0.01              & 3.762 (2.539-4.985), p$<$0.05             \\
    Swin-B                        & 2.788 (1.507-4.069), p$<$0.01               & 2.582 (1.284-3.880), p$<$0.05              & 3.646 (2.006-5.286), p$<$0.001            \\
    MAE 3D                        & 2.755 (1.901-3.609), p$<$0.001              & 2.654 (1.572-3.736), p$<$0.001                & 4.116 (2.593-5.639), p$<$0.05    \\
    MoCo V3                       & 2.704 (1.488-3.920), p$<$0.001             & 2.731 (1.489-3.973), p$<$0.001                 & 3.887 (2.139-5.635), p$<$0.05 \\
    SwinUNETR                     & 2.714 (1.430-3.998), p$<$0.05              & \underline{2.497 (1.443-3.551)}, p$<$0.05              & 3.686 (2.282-5.090), p$<$0.001            \\
    BrainSegF.                    & \underline{2.528 (1.481-3.575)}, p$<$0.001             & 2.573 (1.447-3.699), p$<$0.05              & \underline{3.443 (1.958-4.928)}, p$<$0.01             \\
    Ours                          & \textbf{2.479 (1.480-3.478)}                        & \textbf{2.276 (1.306-3.246)}                        & \textbf{3.417 (1.814-5.020)}                       \\
    \midrule
    $\Delta$ (Swin-B)             & -0.309 (**)                                & -0.306 (*)                                 & -0.229 (***)                              \\ \bottomrule
    \end{tabular}
    }
\end{table*}
    
% \begin{figure}
%     \centering
%     \includegraphics[width=0.5\textwidth]{figures/progression.png}
%     \caption{Progression of Alzheimer's disease(AD) and osteoarthritis(OA), we report the accuracy in the tasks. }
%     \label{fig:progression}
% \end{figure}

\subsection{Cross-sequence Registration}

Accurate registration between MRI sequences with differing contrast settings is essential for multi-parametric analysis, lesion tracking, and surgical planning. This task requires models to extract contrast-invariant and anatomically consistent representations across modalities.

We evaluated the performance of \modelname~on the image registration task using two datasets: the IXI dataset~\cite{dsIXIDataset} for \textit{Independent} validation, and the OASIS 1 dataset~\cite{dsoasis1marcus2007open} for \textit{Held-out} validation. The TransMorph architecture~\cite{chen2022transmorph} was adopted as the registration backbone, with encoder weights initialized either randomly (Swin-B) or using various pre-trained models (MAE 3D, MoCo V3, SwinUNETR, BrainSegF., and \modelname).

As shown in Table~\ref{tab:reg}, Swin-B achieved average Dice scores of 73.74\% on IXI and 83.57\% on OASIS. When initialized with \modelname~pre-trained weights, performance improved to 74.21\% and 86.78\%, corresponding to absolute gains of 0.47\% ($p<0.05$) and 3.22\% ($p<0.001$), respectively. While all pre-trained initializations improved registration accuracy, \modelname~consistently outperformed existing baselines across both datasets, highlighting the robustness and transferability of the learned anatomical representations.

\begin{table}[]
    \centering
    \caption{We report the Dice score for image registration. Noted that all methods performed similarly in \textit{Independent} validation set (IXI), and our model improved over 3\% of Dice score in \textit{Held-out} validation set (OASIS).}
    \label{tab:reg}
        \begin{tabular}{lll}
            \toprule
            Model\textbackslash   Dataset     & IXI~\cite{dsIXIDataset}   & OASIS~\cite{dsoasis1marcus2007open} \\
            \midrule
            TransMorph~\cite{chen2022transmorph} & \underline{73.95 (±1.57)} & \underline{85.59 (±0.66)} \\
            Swin-B                        & 73.74 (±1.29)             & 83.57 (±0.86)             \\
            MAE 3D                        & 73.82 (±2.35)             & 83.27 (±0.80)             \\
            MoCo V3                       & 73.62 (±2.18)             & 84.26 (±0.88)             \\
            SwinUNETR                     & 73.80 (±2.25)             & 85.38 (±0.77)             \\
            BrainSegF.                    & 73.93 (±1.49)             & 84.82 (±0.86)             \\
            Ours                          & \textbf{74.21 (±1.25)}    & \textbf{86.78 (±0.71)}    \\
            \midrule
            $\Delta$ (Swin-B)             & 0.47 (*)                  & 3.22 (***)         \\
            \bottomrule
            \end{tabular}
\end{table}

\subsection{Report Generation}
Automated report generation offers a promising avenue to streamline radiological workflows by producing structured draft reports, thereby reducing dictation time and manual documentation. We evaluated \modelname~in a zero-shot or few-shot setting using a private knee MRI dataset with paired radiology reports, benchmarking the generated outputs against reference reports using standard language generation metrics. This task highlights the feasibility of leveraging a general-purpose visual encoder for vision-to-language transfer in clinical reporting.

\begin{figure}
    \centering
    \includegraphics[width=0.45\textwidth]{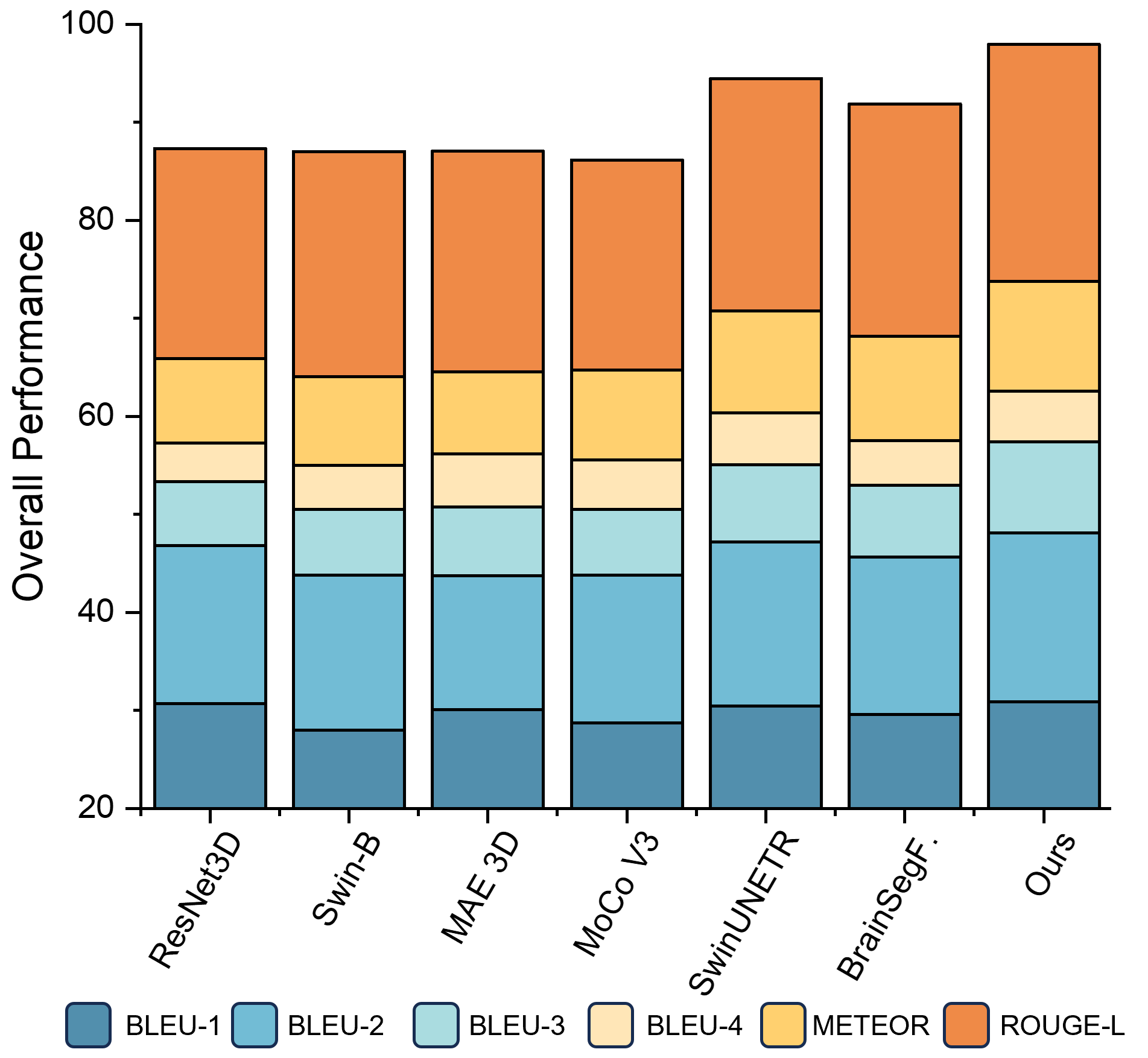}
    \caption{Evaluation of Report Generation Quality on the Private-Knee\_int Dataset. We adapted the R2GenGPT model~\cite{wang2023r2gengpt} to generate both observation and diagnosis sections for MRI reports. Using six common metrics to evaluate the generated reports, our model achieved the best overall performance, demonstrating its superiority as a vision encoder.}
    \label{fig:report_gen}
\end{figure}

\modelname~was used as the vision encoder and integrated with a pre-trained LLaMA decoder~\cite{touvron2023llama} to generate textual reports, thereby assessing the model’s cross-modal understanding capability. Performance was evaluated using three widely adopted natural language generation (NLG) metrics: BLEU (1–4), METEOR, and ROUGE-L. As shown in Fig.~\ref{fig:report_gen} and Supplementary STable 16, \modelname~achieved the highest scores in five of the six metrics, with an average improvement of 2.1\% over the supervised Swin-B baseline. In particular, \modelname~achieved a BLEU-1 score of 30.84, surpassing Swin-B by 2.88 ($p<0.01$). For BLEU-3, which better captures mid-range n-gram coherence, it improved performance by 2.61 ($p<0.01$). Even in BLEU-4, where MAE 3D slightly outperformed other methods with a score of 5.41, \modelname~remained competitive with a score of 5.19, exceeding Swin-B by 0.69 ($p<0.01$). Regarding semantic relevance and fluency, \modelname~also attained the highest scores in METEOR (11.19) and ROUGE-L (24.16), with improvements of 2.15 ($p<0.001$) and 1.17 ($p<0.01$), respectively. These results suggest that MRI-specific pretraining not only enhances discriminative performance but also facilitates rich and coherent language generation in cross-modal tasks. The consistent gains across lexical (BLEU), semantic (METEOR), and structural (ROUGE) metrics validate the generalization capability and vision-language understanding of \modelname.

\subsection{Scalability, Pretraining Strategy, and Adaptation Efficiency}
To evaluate the scalability and transferability of \modelname, we first conducted systematic experiments across multiple downstream tasks, covering sequence identification, age estimation, classification, and segmentation (Supplementary STable 17), under varying model capacities and pretraining dataset sizes (Table~\ref{tab:scal_law}). When increasing the pretraining dataset size from 10k to 53k samples, we observed consistent performance improvements across all tasks except for a margin decrease in disease grading. For example, sequence identification accuracy increased by 1.08\% (from 76.87\% to 77.95\%), and further scaling to 336k samples yielded continued gains (e.g., +0.26\% accuracy), suggesting that data diversity enhances generalization. 
However, increasing model capacity from Swin-B to Swin-H led to diminishing returns. For instance, sequence identification accuracy improved only marginally (+0.04\%, from 78.21\% to 78.25\%), whereas performance on other tasks exhibited fluctuations alongside an overall decreasing trend.
Given the 4× higher computational cost of larger models, we identify Swin-B trained on 336k data as the optimal configuration for balancing efficiency and performance across diverse medical imaging tasks.

To investigate the contribution of our proposed pretraining strategies, we conducted an ablation study focused on three tasks beyond masked image reconstruction (MRI), which has been validated in prior work~\cite{hatamizadeh2021swin}: metadata prediction (Meta), image translation (Tran), and anatomical contrastive learning (Con). The model was pre-trained on the 53k dataset and fine-tuned on the same downstream tasks for scalability evaluation. As reported in Table~\ref{tab:ablation}, metadata prediction yielded substantial gains in sequence identification (+1.17\%), due to its semantic alignment with sequence-aware feature learning. While image translation alone provided limited benefits in abnormality diagnosis (-0.03\%) and lesion segmentation (+0.20\%), it synergized with contrastive learning to boost performance, indicating that combining appearance alignment and anatomical invariance helps the model learn optimal representations. In contrast, image reconstruction showed strong segmentation performance but underperformed in classification, likely due to its intra-sequence reconstruction bias.

\begin{figure}
    \centering
    \includegraphics[width=0.5\textwidth]{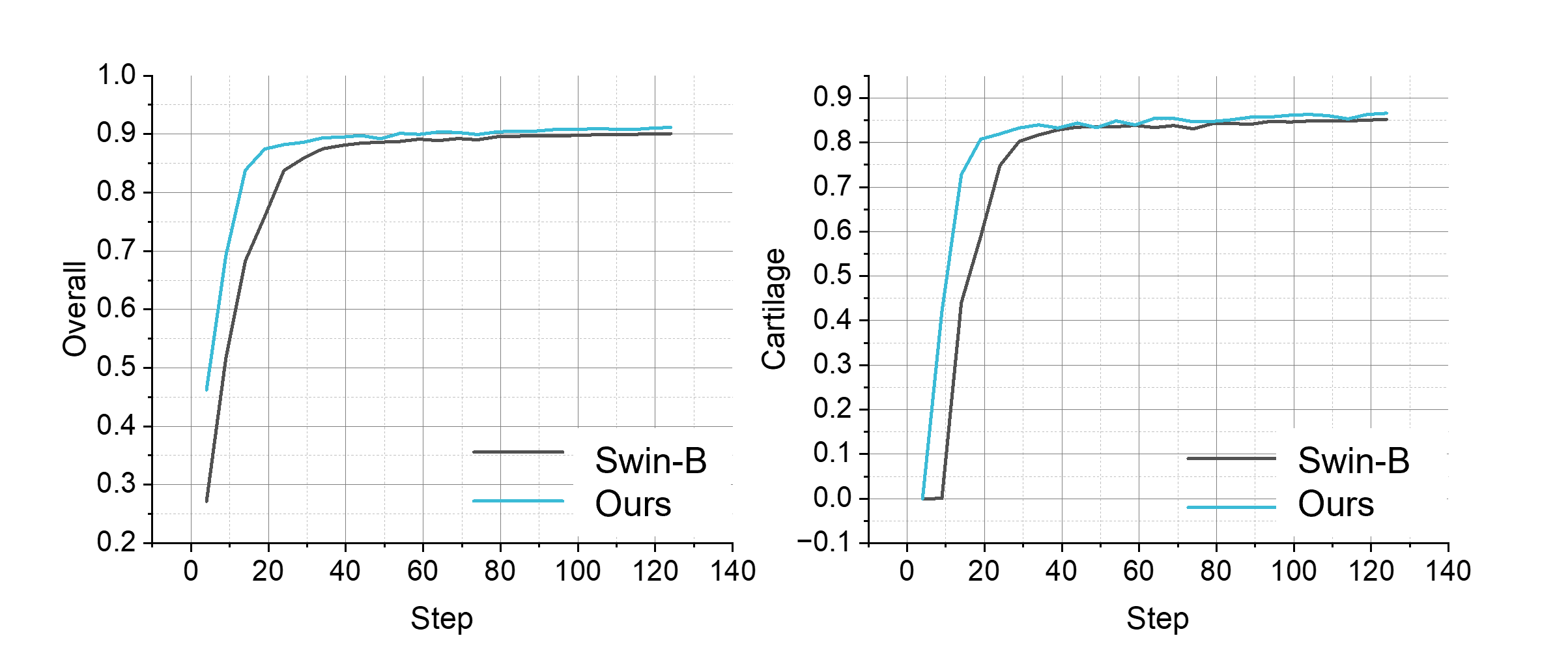}
    \caption{The Dice score curve in fine-tuning stage on OAI-ZIB dataset. We illustrate the curve of overall (left) and cartilage (right) segmentation performance.}
    \label{fig:eff_ft}
\end{figure}

We further assess fine-tuning efficiency using the OAI-ZIB dataset for knee segmentation. As shown in Fig.~\ref{fig:eff_ft}, our model achieved higher initial performance, steeper learning curves, and significantly faster convergence than Swin-ViT trained from scratch, both in overall segmentation and cartilage-specific Dice scores. These findings highlight the benefits of MRI-specific pretraining in reducing the computational cost and data requirement of downstream adaptation. The ability to reach optimal performance with fewer epochs and limited data supports the scalable deployment of \modelname~in real-world clinical scenarios.

% \begin{figure*}
%     \centering
%     \includegraphics[width=0.9\textwidth]{figures/tsne.png}
%     \caption{}
%     \label{fig:reg_vis}
% \end{figure*}

\begin{table*}[]
\centering
\caption{Performance comparison in different scales, with sequence identification, age estimation, classification and segmentation. The numbers in brackets indicate the performance changes from the previous row. The best-performing model is denoted in bold font, and the second-best model is underlined.}
\label{tab:scal_law}
\resizebox{\textwidth}{!}{
\begin{tabular}{lllllll}
    \toprule
Model\textbackslash   Dataset & Sequence   Ide.           & Age Est.                 & Disease Grad.             & Abnormality   Diag.       & Organ   Seg.              & Lesion   Seg.     \\ \midrule
Swin-B (10k)                  & 76.87                     & 2.88                     & 86.25                     & 79.41                     & 89.04                     & 71.21                     \\
Swin-B (53k)                  & 77.95 (↑1.08)             & 2.83 (↓0.05)             & 86.18 (↓0.07)             & 79.69 (↑0.28)             & \textbf{90.62 (↑1.58)}    & 71.55 (↑0.34)             \\
Swin-B (336k)                 & \underline{78.21 (↑0.26)} & \textbf{2.72 (↓0.11)}    & \textbf{86.78 (↑0.60)}    & \textbf{80.05 (↑0.35)}    & 90.42 (↓0.21)             & \textbf{72.09 (↑0.54)}    \\
Swin-L (336k)                 & 78.12 (↓0.09)             & 2.88 (↑0.15)             & \underline{86.53 (↓0.25)} & \underline{79.77 (↓0.28)} & \underline{90.62 (↑0.20)} & \underline{71.59 (↓0.50)} \\
Swin-H (336k)                 & \textbf{78.25 (↑0.13)}    & \underline{2.81 (↓0.06)} & 85.38 (↓1.15)             & 77.46 (↓2.31)             & 90.40 (↓0.22)             & 71.50 (↓0.09)             \\ \bottomrule
\end{tabular}
}
\end{table*}

\begin{table*}[]
\centering
\caption{An ablation study was conducted on a 53k pretraining dataset across four downstream applications. Performance evaluation incorporated accuracy, mean error, and Dice score metrics. The performance changes from the previous row are presented in brackets. Swin-B was adopted as the baseline model to assess the efficacy of our pretraining components, including masked image reconstruction (MIR), metadata prediction (Meta), image translation (Tran), and anatomical-invariant contrastive learning (Con).}
\label{tab:ablation}
\resizebox{\textwidth}{!}{
\begin{tabular}{lllllll}
\toprule
Model\textbackslash   Dataset & Sequence   Ide.   & Age Est.           & Disease Grad.           & Abnormality   Diag.   & Organ   Seg.      & Lesion   Seg.     \\ \midrule
Swin-B                        & 74.40                     & 3.01                  & 84.94                     & 77.78                     & 89.42                     & 70.87                     \\
MIR                           & 75.12 (↑0.72)             & 3.02 (↑0.01)          & 85.04 (↑0.10)             & 77.69 (↓0.09)             & 90.00 (↑0.58)             & 71.04 (↑0.17)             \\
MIR+Meta                      & 76.29 (↑1.17)             & 2.96 (↓0.05)          & 85.20 (↑0.16)             & \underline{78.32 (↑0.63)} & 89.84 (↓0.17)             & \underline{71.09 (↑0.05)} \\
MIR+Meta+Tran                 & \underline{76.69 (↑0.40)} & \underline{2.89 (↓0.07)}& \underline{85.76 (↑0.56)} & 78.29 (↓0.03)             & \underline{90.28 (↑0.45)} & 71.29 (↑0.20)             \\
MIR+Meta+Tran+Con (Ours)      & \textbf{77.95 (↑1.26)}    & \textbf{2.83 (↓0.06)} & \textbf{86.18 (↑0.42)}    & \textbf{79.69 (↑1.40)}    & \textbf{90.62 (↑0.34)}    & \textbf{71.55 (↑0.25)}    \\ \bottomrule

\bottomrule
\end{tabular}
}
\end{table*}

\section{Discussion}

We present \modelname, an MRI-specific foundation model developed to support a broad range of clinical imaging applications. It was pre-trained on \pretrainMRI~multi-sequence MRI volumes from \pretraindataset~datasets, covering major anatomical regions and diverse acquisition protocols. This work marks a substantial step toward building clinically applicable foundation models for medical imaging. While foundation models have transformed natural image and language tasks, their adoption in volumetric medical imaging, especially MRI, remains limited. By constructing the large-scale MRI-specific foundation model pre-trained on over 330,000 sequences, we provide compelling evidence that large-scale representation learning significantly improves accuracy, robustness, and transferability across diverse MRI tasks. Evaluation across \downstreamdataset~datasets demonstrates state-of-the-art performance in versatile applications, underscoring the model’s potential to support real-world clinical workflows. \modelname~effectively captures shared anatomical patterns and sequence variations, enabling it to bridge heterogeneity across scanner protocols, institutional settings, and patient populations.

A key innovation of \modelname~is its pretraining framework that disentangles anatomical-invariant and sequence-specific representations via four complementary self-supervised tasks, enabling robust feature transfer across anatomical regions and imaging protocols, reducing reliance on organ-specific annotations and supporting generalization under domain shifts.
Our results demonstrate that \modelname~achieves state-of-the-art Dice scores across both organ and lesion segmentation tasks, reflecting its capacity to capture fine-grained anatomical details essential for surgical planning and radiological assessment. 
In disease grading and abnormality detection, the model leverages semantically rich and anatomically grounded representations to achieve high classification accuracy across varied imaging protocols and clinical centers. 
Notably, \modelname~generalizes robustly in zero-shot and external validation settings, even under substantial distribution shifts introduced by differences in scanner vendors, acquisition parameters, and patient populations. 
In addition to classification and segmentation, \modelname~also exhibits promising capabilities for supporting regression and vision-language generation tasks, with potential to produce clinically relevant outputs such as continuous severity estimations and structured radiology reports. These capabilities are particularly valuable for quantifying subtle imaging biomarkers, supporting early disease monitoring, and promoting standardized documentation in longitudinal and multi-site clinical workflows.
Moreover, \modelname~exhibits remarkable efficiency in downstream adaptation. Compared to models trained from scratch, it converges significantly faster and achieves better performance with limited labeled data, underscoring its label and data efficiency. This is particularly important for real-world medical applications, where data annotation is often expensive and time-consuming.

Despite its strengths, developing a robust foundation model for MRI presents several key challenges. For example, while the pretraining dataset demonstrates considerable diversity, it exhibits significant anatomical imbalance: over half of the volumes originate from knee MRI scans. This imbalance risks undermining the model's generalizability to underrepresented organs and anatomical regions. Although we incorporated multi-region data to alleviate this issue, performance improvements remain disproportionately concentrated in knee-related tasks. Our scaling experiments further reveal that data volume and diversity, rather than anatomical bias, serve as the primary drivers of performance enhancements, consistent with established observations from scaling laws.
We recognize that developing a dedicated data engine or standardized curation workflow will be critical to systematically addressing this imbalance. Such infrastructure would enable more equitable performance across diverse anatomical regions while enhancing the feasibility of building even larger, more robust models.

In this work, we have compared several strong baseline models (i.e., nnUNet), and our experiments were conducted using a fixed backbone structure (Swin-ViT). While our model may not surpass specialized models in certain tasks, the pre-training paradigm we propose can be applied to various model structures and demonstrates strong scalability. we observe diminishing returns when scaling model capacity beyond Swin-B, while data scaling continues to yield performance gains. This suggests that data diversity, rather than model size, is the primary driver of generalization in current MRI applications, an insight with practical implications for efficient foundation model design.

Future work will explore improved representation learning strategies to mitigate these limitations. While data diversity alleviates sequence variability, it does not fully resolve domain shift issues. We therefore plan to investigate domain alignment and normalization strategies to further improve robustness. We also aim to expand \modelname~into the vision-language domain, jointly pretraining on MRI images and paired clinical reports to enhance semantic understanding and interpretability. Unifying visual and textual supervision may support broader deployment and foster explainable AI in clinical radiology.

In summary, \modelname~establishes a scalable and effective foundation for MRI-based medical AI. Through anatomy-aware and sequence-sensitive representation learning at scale, the model delivers strong generalization, robust downstream performance, and efficient fine-tuning across diverse clinical tasks. We hope this work encourages broader exploration of generalizable and semantically aligned foundation models for real-world medical imaging applications.
\section{Methods}
\subsection{Data Acquisition and Preprocessing}
This study received ethical approval from the Human and Artefacts Research
Ethics Committee (HAREC) at the Hong Kong University of Science and Technology (No.HREP-2025-0230). 

We curate \AlldsNumber~datasets from diverse public and private sources, establishing a pretraining cohort that includes \SeqNumber~multi-sequence MRI. The data spans diverse anatomical regions including head (brain, neck), chest (breast), abdomen (liver, pancreas, rectum, bladder), reproductive system (uterus, prostate), and lower body (knee), with $7$ common sequences: T1W, T2W, proton density weighted (PDW), dynamic contrast-enhanced (DCE), short tau inversion recovery (STIR), diffusion-weighted imaging (DWI), and fluid-attenuated inversion recovery (FLAIR). These scans, acquired across multiple centers with heterogeneous imaging protocols and MRI vendors, reflect substantial diversity in acquisition conditions and clinical cohorts. 

In general, the 336k dataset includes \pretrainpublic~public datasets and \pretrainprivate~private datasets, as shown in Fig.~\ref{fig:pipeline} (e). Our private data was curated from \PrivCenter~ different centers, with four centers dedicated to external testing. The largest component of our private data is Private-Knee\_int, comprising 10,000 cases and 50,000 volumes. The remaining private datasets are categorized by disease type based on clinical visits, encompassing 9 distinct cancer types. The 10k and 53k datasets are subsets of the 336k dataset.
Notably, we adopt the fold list for the BraTS dataset provided by BrainSegFounder~\cite{cox2024brainsegfounder} to prevent data leakage.
We recorded the detailed information of all private and public dataset in the supplementary file.

After data collection, we performed a series of preprocessing steps. First, the raw MRI volumes were resampled to an isotropic spacing of $1.0 \times 1.0 \times 1.0$. A random crop of size $96 \times 96 \times 96$ was then applied to extract sub-volumes. To simulate variations in MRI acquisition planes, the images were randomly reoriented. Subsequently, 30\% of the regions of interest (ROIs) were randomly masked, following the strategy in~\cite{hatamizadeh2021swin}, to create partially occluded inputs. Additionally, three essential scanning parameters, i.e., repetition time (TR), echo time (TE), and flip angle (FA), were extracted for use in pretext tasks.

\begin{figure*}
    \centering
    \includegraphics[width=\linewidth]{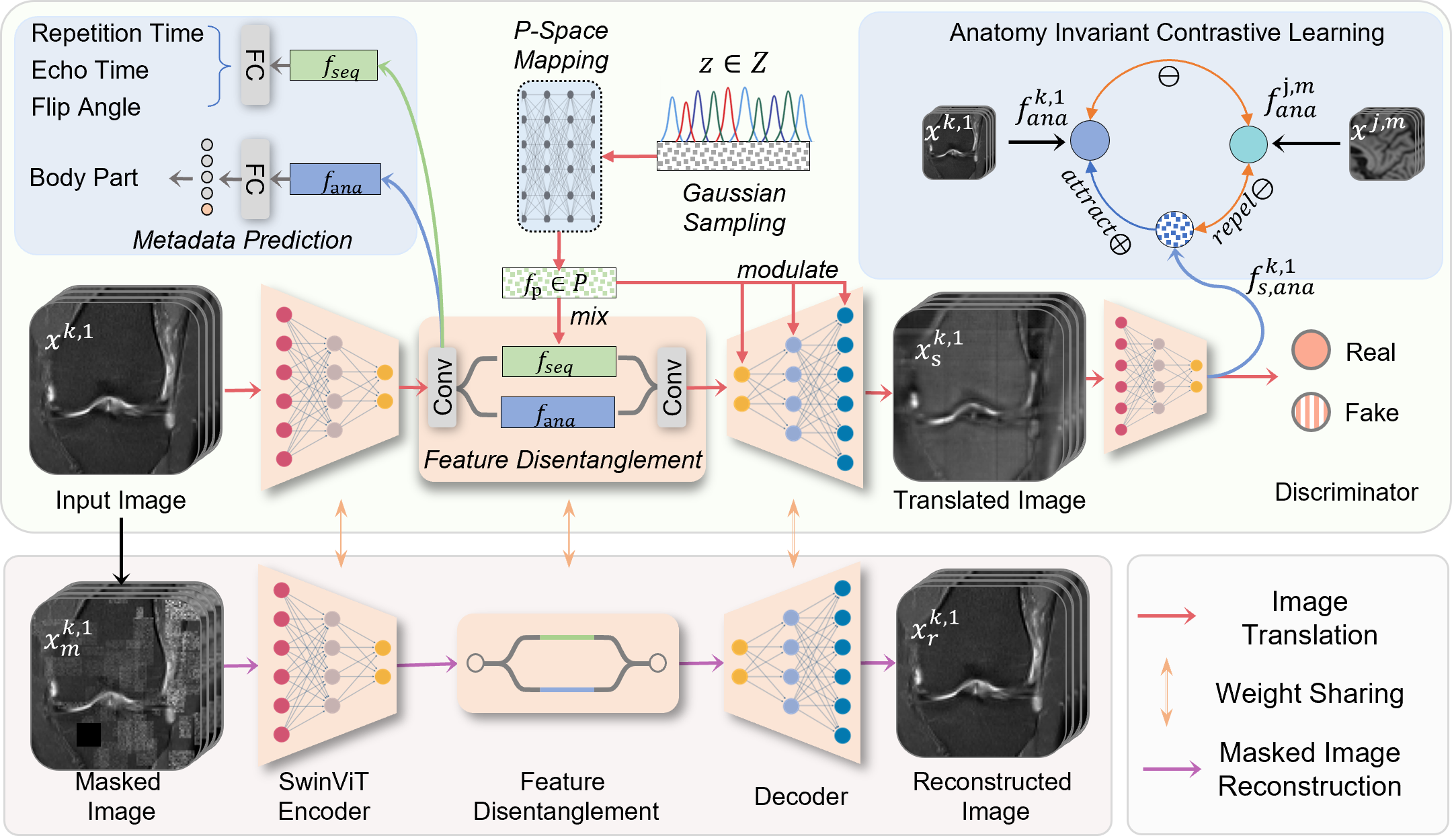}
    \caption{ Overview of proposed pretraining process. We adopt Swin transformer (SwinViT) as the encoder architecture, whose output features are disentangled into anatomical features and sequence features (denoted as blue and green in the figure). To facilitate effective pretraining, we employ two generative tasks: masked image reconstruction and GAN-based image translation. Additionally, we incorporate two latent-space regulations: metadata prediction and anatomy-invariant contrastive learning.}
    \label{fig:overview}
\end{figure*}

\subsection{Network Architecture and Pretext Tasks for Pretraining}
The proposed architecture comprises a Swin Transformer~\cite{hatamizadeh2021swin} backbone for hierarchical feature extraction and long-range dependency modeling, coupled with a dual-branch disentanglement module, as shown in Fig.~\ref{fig:overview}. Features extracted by the encoder are partitioned into two subspaces: \textit{anatomical features} that are invariant across sequences, and \textit{sequence-specific features} that capture acquisition-dependent variations (e.g., T1/T2 weighting, proton density). These features form the basis for the subsequent four pretext tasks, including pixel-level masked image reconstruction, cross-sequence translation, metadata prediction and anatomy-invariant contrastive learning, which are jointly trained in an end-to-end manner.

\noindent\textbf{Masked image reconstruction.} It aims to recover missing or occluded regions in input images, and is widely employed in unsupervised representation learning to encourage spatial awareness and contextual understanding. Following prior work ~\cite{tang2022self}, we adopt a volumetric masked reconstruction strategy tailored for 3D MRI data. Specifically, a contiguous cubic region occupying a volume ratio of 
$s$ is randomly masked from the input MRI volume $x^{i,n}$, simulating partial observability during pretraining.

To reconstruct the occluded content, we append a transposed convolutional layer as a lightweight reconstruction head to the encoder output. This head generates a reconstructed image volume, which is compared against the original unmasked input using an L1 loss function: $\mathcal{L}_{Recon} =||x^{i,n} - \hat{x}^{km,}_i|| _1$. This reconstruction objective compels the encoder to learn spatially coherent and anatomically consistent feature representations, even in the presence of incomplete visual information. By restoring structure from partial data, the model develops a stronger inductive bias toward anatomical integrity, thereby enhancing its utility for downstream tasks such as segmentation and diagnosis.

\noindent\textbf{P-space guided image translation.}
The image translation task focuses on synthesizing anatomically aligned images with alternative contrast properties.

MRI inherently exhibits variability due to differences in acquisition parameters~\cite{marzi2024efficacy}. To systematically model this heterogeneity, we introduce a latent parameter space (referred to as \textit{P-space}) that captures the distribution over all plausible combinations of acquisition protocols. This space enables the model to learn compact representations of sequence-dependent imaging characteristics, facilitating structured decomposition of modality-specific features.

Inspired by generative disentanglement strategies in the StyleGAN framework~\cite{karras2019style}, we design a P-space-guided translation pipeline that explicitly decouples anatomical content from contrast variations. As illustrated in the top half of Fig.~\ref{fig:overview}, the pipeline begins by disentangling the encoder-derived deep features into two orthogonal components: an anatomical representation $f_{\text{ana}}$, shared across sequences, and a sequence-specific representation $f_{\text{seq}}$.

To simulate contrast variation, a latent vector $z$ is sampled from a standard Gaussian space $Z$, and mapped via a multilayer perceptron (MLP) into a parameterized feature vector $f_p \in \text{P-space}$. This vector $f_p$ is then fused with the sequence-specific feature $f_{\text{seq}}$, and the combined representation is subsequently integrated with $f_{\text{ana}}$ through a convolutional layer to preserve anatomical integrity. The aggregated feature is passed to a decoder, which synthesizes an output image that maintains the structural content of the input while exhibiting the contrast style defined by the sampled parameters.

To further enforce realism and parameter fidelity, the synthetic image is re-encoded, and a latent-space discriminator evaluates whether the resulting feature distribution is indistinguishable from that of real images. The generator and discriminator are jointly optimized using an adversarial loss, ensuring that the sampled feature $f_p$ spans the full spectrum of clinically plausible contrast variations. This approach enables anatomically faithful, parameter-controlled image translation and supports robust generalization across MRI protocols.

\noindent\textbf{Anatomy invariant contrastive learning.} Contrastive learning enhances representation learning by maximizing mutual information between positive pairs while minimizing similarity to negative pairs, offering a versatile framework for diverse data types. However, its efficacy critically depends on the strategic selection of positive and negative pairs. While conventional methods generate positive pairs via augmentations (e.g., cropping, rotation) from a single data instance and treat all other instances as negatives, this approach is suboptimal for medical imaging. Unlike natural images, medical data contains fine-grained anatomical structures that are sensitive to intensity variations introduced by standard augmentations (e.g., contrast adjustments), risking corruption of biologically meaningful features. To address this, we propose leveraging inter-sequence correlations, where anatomical consistency is preserved between an original image and its synthetic counterpart $f^{k,1}_s$.
By attracting the synthetic anatomical features $f^{k,1}_{s,ana}$ and repelling the others, the model can capture and disentangle anatomical features from intensity contrast changes.

We use $f^{n,i}_{\text{ana}}$ to denote the anatomical feature extracted from the $i$-th sequence of the $n$-th subject, and $f^{n,i}_{s,\text{ana}}$ to represent the corresponding feature from the translated synthetic image. Following the InfoNCE formulation~\cite{chen2020simple}, we define the anatomy-invariant contrastive loss as:

\begin{equation}
\mathcal{L}_{Con} = -log\frac{exp(f^{n,i}_{ana}\cdot f^{n,i}_{s,ana}/\tau)}{\sum_{k=0}^{K}{exp(f^{n,i}_{ana}\cdot f^{m,k}_{ana}/\tau)}}
\end{equation}

\noindent
where $\tau$ is a temperature scaling factor, and the denominator includes negative samples from other subjects or sequences within the batch. This formulation encourages anatomical features to be invariant across varying contrast domains while maintaining discriminability between anatomically distinct regions.

\noindent\textbf{MRI metadata prediction}
To provide structured supervision that guides representation disentanglement, we incorporate a multi-task auxiliary objective during pretraining, targeting key elements of MRI metadata. Specifically, we introduce two predictive tasks: scanning parameter regression and anatomical region classification.

For scanning parameter regression, the model predicts acquisition-specific parameters, repetition time (TR), echo time (TE), and flip angle (FA), using the disentangled sequence-specific feature representation $f_{\text{seq}}$. These predictions are optimized via a mean absolute error (MAE) loss. While these parameters do not fully determine image contrast, they serve as proxies for contrast modulation, encouraging the model to capture sequence-related physical properties that influence image appearance.

In parallel, the anatomical representation $f_{\text{ana}}$ is used to classify the scanned body region (e.g., brain, abdomen, pelvis), providing semantic supervision over spatial structure. This body part classification task enhances the anatomical specificity of learned features, reinforcing structural consistency across sequences.

Together, these metadata prediction tasks promote the development of physiologically and physically grounded representations, improving the model’s ability to disentangle anatomical and contrast-specific information. This design facilitates generalization across imaging protocols and anatomical regions in downstream applications.

\subsection{Adaptation to Downstream Tasks}
This study proposed a pre-trained vision encoder with Swin Transformer backbone, which can be easily adapted to multiple downstream tasks, see Fig.~\ref{fig:overview}: The downstream task can be categorized into four types according to the setting of input and output. For segmentation that output pixel-wise prediction, we incorporate a SwinUNETR~\cite{hatamizadeh2021swin} structure. For classification and regression tasks, we append MLP to the vision encoder. To achieve image registration that align the anatomical structure across multiple sequences, we leverage the TransMorph architecture. For report generation, we introduce large language model (LLM) in R2GenGPT~\cite{wang2023r2gengpt}.

We adapted the pre-trained Swin Transformer encoder to a range of downstream medical tasks by removing the pretext decoders and utilizing the encoder to extract high-level feature representations from input MRI volumes. Task-specific heads were attached depending on the target application. For segmentation tasks, the encoder was integrated into a SwinUNETR architecture~\cite{hatamizadeh2021swin}, where a convolutional decoder generated voxel-wise predictions. During training, the input for segmentation was cropped into $96 \times 96 \times 96$ patches, and sliding-window inference was employed during validation. For datasets with pre-registered sequences (e.g., BraTS), the data were concatenated to form a multi-channel input. For classification and regression tasks, each different sequence was passed through a dedicated encoder, as shown in Fig.~\ref{fig:classification}(a). The extracted features were concatenated and fed through a multilayer perceptron (MLP) to produce scalar or categorical outputs. In image registration, the encoder was embedded into the TransMorph framework to estimate deformation fields that align anatomical structures across sequences. For report generation, the encoder was connected to a pre-trained LLaMA encoder following the R2GenGPT pipeline~\cite{wang2023r2gengpt}, enabling synthesis of structured clinical descriptions from image embeddings.

For each downstream task, model weights were fine-tuned using labeled data. The training hyperparameters, including batch size, optimizer, and learning rate, were adjusted according to dataset conditions. However, these hyperparameters were kept consistent across comparison models for the same task to ensure fairness in evaluation. Unless otherwise specified, training runs for 200 epochs, and the best-performing model, defined by the highest validation metric (e.g., Dice score for segmentation, accuracy for classification), is saved as the final checkpoint. Data augmentations applied during training include isotropic resizing, random horizontal and vertical flipping, random rotation, intensity normalization, and contrast jittering to enhance model generalizability.

\subsection{Evaluation Metrics}
We report accuracy (ACC) and its 95\% confidence interval (CI) for classification tasks, including abnormality diagnosis, disease grading, sequence identification, and progression prediction. Additionally, for binary abnormality classification tasks, we report the area under the receiver operating characteristic curve (AUC-ROC) to enable evaluation independent of decision threshold selection. For segmentation and registration tasks, we report the Dice score and its 95\% confidence interval. For age prediction tasks, we utilize the mean absolute error (MAE) to quantify the discrepancy between predicted and actual ages. In the report generation task, we compute BLEU (1-4)~\cite{papineni2002bleu} by comparing n-gram overlaps between generated reports and ground truth references. Furthermore, we also report METEOR~\cite{denkowski2011meteor} and ROUGE-L~\cite{lin2004rouge} to assess the quality of the generated reports.

\subsection{Computational Resources}

\noindent All self-supervised pretraining was conducted on an NVIDIA SuperPOD cluster equipped with 8 NVIDIA H800 GPUs. The framework was implemented using PyTorch (v2.0)~\cite{Ansel_PyTorch_2_Faster_2024} and MONAI (v1.2)~\cite{Cardoso_MONAI_An_open-source_2022}, which provide modular support for medical image processing and model development. The model was pre-trained using four synergistic pretext tasks, each associated with a self-supervised loss function, to learn robust and generalizable representations from multi-sequence MRI data.

\subsection{Statistical Analysis}
The statistical analysis is conducted using the rpy2 package~\cite{rpy2} and SciPy~\cite{2020SciPy-NMeth} in Python. 
Unless specified otherwise, the Delong test~\cite{delong1988comparing} is employed to assess the significance of differences in AUC-ROC, while the Wilcoxon test is used for ACC and DSC using bootstrapping. 
We compute the confidence interval via the bootstrapping method with 1000 iterations. 
The significance level is defined as p$\leq$0.05 for statistical significance, and we also report significance at level p$\leq$0.01 and p$\leq$0.001.

\section{Data Availability}
The detailed information of all datasets used in this study can be found in the Supplementary file. 
Public datasets can be requested through their respective sources, and we provide direct links to enable researchers to access the relevant data for verification or extended analytical investigations. 
All private datasets are supervised by the corresponding institutions. The data was used with institutional permission, as approved by a review board. Due to data restrictions applied in this study, the data is currently unavailable to the public.
% Please email the corresponding author with all requests for academic use of raw and processed data. 
% The requirements will be evaluated with regard to institutional policies, and data can only be shared for non-commercial
% academic usage with a formal material transfer agreement.
Source data are provided with this paper.

\section{Acknowledgements}
This work was supported by the Health and Medical Research Fund (Ref:20211021), the Health Bureau, The Government of the Hong Kong Special Administrative Region, Hong Kong Innovation and Technology Commission (Project No. GHP/006/22GD, MHP/002/22, and ITCPD/17-9), and National Key R\&D Program of China (Project No. 2023YFE0204000). This work was also supported in part by the National Natural Science Foundation of China under grant 62402458.

We thank all the public dataset providers in this study whose contributions have significantly advanced interdisciplinary research. 
\begin{itemize}
\item ADNI \cite{dsADNImueller2005alzheimer}: 
Data used in preparation of this article were obtained from the Alzheimer's Disease
Neuroimaging Initiative (ADNI) database (adni.loni.usc.edu). As such, the investigators
within the ADNI contributed to the design and implementation of ADNI and/or provided data
but did not participate in analysis or writing of this report. A complete listing of ADNI
investigators can be found at:
\url{http://adni.loni.usc.edu/wp-content/uploads/how_to_apply/ADNI\_Acknowledgement\_List.pdf}

\item fastMRI \cite{dsfastMRIknoll2020fastmri}: 
Data used in the preparation of this article were obtained from the NYU fastMRI Initiative database (fastmri.med.nyu.edu)

\item PPMI \cite{dsPPMImarek2018parkinson}:
Data used in the preparation of this article was obtained from the Parkinson's Progression Markers Initiative (PPMI) database 
(\url{https://www.ppmi-info.org/access-dataspecimens/download-data}), RRID:SCR\_006431. 
For up-to-date information on the study, visit \url{https://www.ppmi-info.org}. 
PPMI – a public-private partnership – is funded by the Michael J. Fox Foundation for Parkinson's Research, and funding partners; included in \url{https://www.ppmi-info.org/about-ppmi/who-we-are/study-sponsors} 
\end{itemize}
%\input{sections/CodeAvailability}
% \input{suppl}

% \begin{appendices}

% \section{No appendix}\label{secA1}




% \end{appendices}


\bibliographystyle{unsrt}
\bibliography{sn-bibliography}% common bib file

\end{document}

\title[Article Title]{Supplementary File}

\maketitle

\section{Data Details}

We collected a total of \alldataset~datasets in this study, comprising 34 from public sources and 30 from private repositories. As presented in STable~\ref{tab:pub_ds_link}, we provide an overview of all public datasets, accompanied by their respective data acquisition links. 
The private data used in this study was collected from \PrivCenter~clinical centers (denoted as A to O in STable~\ref{tab:priv_ds_center}), covering 10 major diseases, including Rectal Cancer (REC), Cervical Cancer (CERV), Head and Neck Squamous Cell Carcinoma (HNSC), Pancreatic Adenocarcinoma (PAAD), Breast Cancer (BRCA), Endometrial Carcinoma (UCEC), Hepatocellular Carcinoma (HCC), Cholangiocarcinoma (CHOL), Bladder Cancer (BLCA), and knee abnormality (KNEE). 
Each dataset is named using the prefix \textit{Private-} followed by the disease abbreviation and the center label, indicating that it is a non-public dataset specifically collected for this study. For example, \textit{Private-HNSC\_A} refers to Head and Neck Squamous Cell Carcinoma cancer MRI data collected from Center A. For convenience, we use \textit{Private-HNSC} to denote the aggregated dataset comprising rectal cancer MRI scans from all participating centers (i.e., A, C, and I). Notably, the five private knee datasets are handled differently: data from Center K (i.e., \textit{Private-Knee\_K}) is designated as \textit{Private-Knee\_int}, while data from Centers L, M, N, and O (i.e., \textit{Private-Knee\_L}, \textit{Private-Knee\_M}, \textit{Private-Knee\_N}, and \textit{Private-Knee\_O}) are merged into a single dataset named \textit{Private-Knee\_ext}. 

In STable~\ref{tab:pretraining}, we present the detailed components of our pretraining data. We include \pretrainpublic~public datasets as well as \pretrainprivate~private dataset (i.e., 10 combined datasets), spanning human body parts from the brain to the knee. To explore different pretraining scales, we prepared datasets with 10k, 53k, and 336k volumes, where the 10k and 53k datasets are subsets reduced from the 336k dataset. We also report the case numbers for better understanding, with each case containing multiple MRI sequences.

\begin{table*}[htp]
\centering
\caption{The publicly available dataset utilised in this study is detailed below. The download link or access website is provided in the second column.}
\label{tab:pub_ds_link}
\begin{tabular}{ll}
        \toprule
        \textbf{Dataset}  & \textbf{Link}                                                                       \\
        \midrule
        ACDC~\cite{dsACDCbernard2018deep}              & https://www.kaggle.com/datasets/anhoangvo/acdc-dataset                            \\
        ADNI~\cite{dsADNImueller2005alzheimer}              & https://adni.loni.usc.edu/data-samples/adni-data/neuroimaging/mri                 \\
        AMOS~\cite{dsAMOSji2022amos}              & https://amos22.grand-challenge.org                                                \\
        ATLAS~\cite{dsATLASquinton2023tumour}             & https://atlas-challenge.u-bourgogne.fr/dataset                                    \\
        BraTS 2021~\cite{dsBraTSbaid2021rsna}        & https://www.cancerimagingarchive.net/analysis-result/rsna-asnr-miccai-brats-2021  \\
        CARE-Liver~\cite{dscare3wu2022meru}        & https://zmic.org.cn/care\_2025/track4                                             \\
        CHAOS~\cite{dsCHAOS2021}             & https://chaos.grand-challenge.org/Publications                                    \\
        DUKE Breast~\cite{dsDUKEBreastsaha2018machine}       & https://wiki.cancerimagingarchive.net/pages/viewpage.action?pageId=70226903       \\
        DUKE Liver~\cite{dsDUKELivermacdonald2023duke}        & https://zenodo.org/records/7774565                                                \\
        EMIDEC~\cite{dsEMIDEClalande2020emidec}              & https://emidec.com/dataset                                                        \\
        fastMRI~\cite{dsfastMRIknoll2020fastmri}           & https://fastmri.med.nyu.edu                                                       \\
        HaN-Seg~\cite{dsHaNpodobnik2023han}           & https://zenodo.org/records/7442914                                                \\
        IXI~\cite{dsIXIDataset}               & https://brain-development.org/ixi-dataset                                         \\
        LLD-MMRI~\cite{dsLLDlou2025sdr}          & https://github.com/LMMMEng/LLD-MMRI-Dataset                                       \\
        MAMA-MIA~\cite{dsmamamiagarrucho2025}          & https://github.com/LidiaGarrucho/MAMA-MIA                                         \\
        MM-WHS~\cite{dsMMWHSzhuang2018multivariate}            & https://zmiclab.github.io/zxh/0/mmwhs                                             \\
        MRNet~\cite{dsMRNetbien2018deep}             & https://stanfordmlgroup.github.io/competitions/mrnet                              \\
        MSD~\cite{dsMSDantonelli2022medical}               & http://medicaldecathlon.com/dataaws                                               \\
        OAI~\cite{dsOAInevitt2006osteoarthritis}               & https://nda.nih.gov/oai                                                           \\
        OAI-ZIB~\cite{dsOAIZIBambellan2019automated}           & https://gitlab.com/vvr/OActive/osteoarthritis\_initiative\_zib\_dataset           \\
        OASIS~\cite{dsOASISlamontagne2019oasis}             & https://sites.wustl.edu/oasisbrains/home                                          \\
        openBHB~\cite{dsOpenBHBdufumier2022openbhb}           & https://baobablab.github.io/bhb/dataset                                           \\
        PanSegData~\cite{dspansegzhang2025large}        & https://osf.io/kysnj                                                              \\
        PI-CAI~\cite{dsPICAIsaha2024artificial}            & https://pi-cai.grand-challenge.org                                                \\
        PPMI~\cite{dsPPMImarek2018parkinson}              & https://www.ppmi-info.org/access-data-specimens/download-data                     \\
        PROMISE12~\cite{dsPROMISE12litjens2014evaluation}         & https://promise12.grand-challenge.org/Home                                        \\
        Prostate158~\cite{dsProstate158adams2022prostate158}       & https://github.com/kbressem/prostate158                                           \\
        SKM-TEA~\cite{dsSKMTEAdesai2022skm}           & https://doi.org/10.71718/2ghb-nv62                                                \\
        SPIDER~\cite{dsSPIDERvan2024lumbar}            & https://spider.grand-challenge.org/data                                           \\
        Total Segmentator~\cite{dsTotalSegd2024totalsegmentator} & https://zenodo.org/records/14710732                                               \\
        \bottomrule
\end{tabular}
\end{table*}

\clearpage

\begin{table*}[htp]
\centering
\caption{Distribution of private MRI datasets across diseases and medical centers. Rows represent diseases, and columns correspond to medical centers (denoted A to O). Each non-empty cell indicates the number of MRI scans collected for that disease\_center pair. The corresponding dataset is referred to by concatenating the disease name and center label (e.g., \textit{Private-HNSC\_A} refers to the head and neck squamous cell carcinoma cancer (HNSC) MRI data collected from Center A. \textit{Private-HNSC} denotes the aggregated dataset comprising HNSC MRI scans from all participating centers (i.e., A, C, and I).}
\label{tab:priv_ds_center}
\resizebox{0.9\textwidth}{!}{
    \begin{tabular}{@{}lllllllllllllllll@{}}
    \toprule
    Disease\textbackslash{}Center & A     & B  & C   & D   & E   & F  & G   & H   & I   & J     & K      & L   & M  & N  & O  & Total  \\ \midrule
    REC                   & 456   &    &     &     &     &    &     &     &     &       &        &     &    &    &    & 456    \\
    CERV                  & 1,042 & 99 & 850 & 93  & 99  & 83 & 118 & 158 &     &       &        &     &    &    &    & 2,542  \\
    HNSC                  & 1,408 &    & 210 &     &     &    &     &     & 449 &       &        &     &    &    &    & 2,067  \\
    PAAD                  & 179   &    &     &     &     &    &     &     &     &       &        &     &    &    &    & 179    \\
    BRCA                  & 1,117 &    &     &     & 235 &    &     & 313 & 182 & 1,166 &        &     &    &    &    & 3,013  \\
    UCEC                  & 526   &    & 583 & 231 &     &    &     & 250 &     &       &        &     &    &    &    & 1,590  \\
    HCC                   & 227   &    &     &     &     &    &     &     &     &       &        &     &    &    &    & 227    \\
    CHOL                  & 206   &    &     &     &     &    &     &     &     &       &        &     &    &    &    & 206    \\
    BLCA                  & 163   &    &     &     &     &    &     &     &     &       &        &     &    &    &    & 163    \\

    KNEE             &       &    &     &     &     &    &     &     &     &       &   10,000      & 500 & 37 & 64 & 44 & 10,645    \\ \bottomrule
    \end{tabular}
}
\end{table*}

\begin{table*}[htp]
\centering
\caption{Pretraining data with multiple body parts and scales. We incorporate 8 public dataset and 10 combined private datasets (26 distinct datasets) to form our pretraining cohort, which includes multiple body parts and common MRI sequences. We prepare three scales of pretraining sizes, namely 10k, 53k, and 336k. The number of cases and MRI sequences are indicated in the Case and Vol columns, respectively. The second column specifies the major body parts of each dataset.}
\label{tab:pretraining}
\begin{tabular}{llllllll}
\toprule
Dataset          & Part     & \multicolumn{2}{l}{10K} & \multicolumn{2}{l}{53K} & \multicolumn{2}{l}{336K} \\ \midrule
                 &          & Case       & Vol        & Case       & Vol        & Case       & Vol         \\
OAI              & Knee     & 500        & 500        & 1,000      & 1,000      & 4,000      & 60,013      \\
OASIS 3          & Brain    & 500        & 500        & 1,000      & 2,000      & 1,000      & 6,814       \\
ADNI             & Brain    & 500        & 500        & 1,000      & 3,000      & 1,000      & 3,617       \\
PI-CAI           & Prostate & 500        & 500        & 500        & 500        & 1,000      & 5,000       \\
fastMRI-Prostate & Prostate & 312        & 312        & 312        & 312        & 312        & 1,247       \\
fastMRI-Brain    & Brain    & 500        & 500        & 2,000      & 2,000      & 10,059     & 23,135      \\
fastMRI-Knee     & Knee     & 500        & 500        & 1,000      & 2,000      & 9,289      & 47,463      \\
fastMRI-Breast   & Breast   & 600        & 600        & 600        & 600        & 600        & 600         \\
Private-Knee\_int & Knee     & 300        & 900        & 1,000      & 2,000      & 10,000     & 50,000      \\
Private-BRCA     & Breast   & 300        & 900        & 2,000      & 6,000      & 3,013      & 55,807      \\
Private-REC      & Rectum   & 300        & 900        & 456        & 6,807      & 456        & 6,807       \\
Private-CERV     & Uterus   & 150        & 450        & 1,000      & 3,000      & 2,542      & 18,558      \\
Private-UCEC     & Uterus   & 150        & 450        & 1,000      & 3,000      & 1,590      & 19,467      \\
Private-HNSC     & Neck     & 300        & 900        & 2,067      & 6,231      & 2,067      & 22,586      \\
Private-PAAD     & Pancreas & 179        & 895        & 179        & 2,994      & 179        & 2,994       \\
Private-HCC      & Liver    & 227        & 454        & 227        & 4,369      & 227        & 4,369       \\
Private-CHOL     & Liver    & 206        & 412        & 206        & 3,934      & 206        & 3,934       \\
Private-BLCA     & Bladder  & 163        & 618        & 163        & 4,065      & 163        & 4,065       \\
Total            &          & 6,187      & 10,791     & 16,710     & 53,812     & 47,703     & 336,476     \\ \bottomrule
\end{tabular}
\end{table*}

\clearpage

\section{Downstream Evaluation}

Our evaluation incorporates \FTtasknumber~downstream tasks, with detailed characteristics of each task summarized in STable~\ref{tab:downstream_split}. 
This includes dataset sources, evaluation splits, case distributions across training/validation/test sets, and anatomical regions. We fine-tune our model on the training split, select the epoch based on the maximum validation performance, and then report the performance on the testing split.  
Notably, some datasets are utilized for multiple subtasks through different processing approaches.

To ensure consistent evaluation, we adhere to the following data splitting protocols:
1) Official splits preserved: For datasets with predefined training, validation, and test partitions, we maintain the original distribution.

2) Unlabeled test sets: When test labels are unavailable, we repurpose the original validation set as the test set and allocate 20\% of the training data for validation.

3) Undivided datasets: For datasets without predefined splits, we partition the data into training, validation, and test sets using a 7:1:2 ratio, respectively.

This stratified approach ensures proper separation of evaluation data while respecting dataset-specific characteristics. All splits were implemented before any model training or parameter optimization to prevent data leakage.
In the following section, we will demonstrate the detailed performance of our benchmarking evaluation. Additionally, we will describe some special datasets and evaluation settings to facilitate a better understanding.


\begin{sidewaystable}[ht]
\centering
\caption{All downstream tasks, their corresponding dataset, evaluation group and the case number in each split.}
\label{tab:downstream_split}
\begin{tabular}{llllllll}
\toprule
Name              & Task                           & Subtask                             & Part       & Group       & Train & Val & Test \\
\midrule
LLD-MMRI          & Abnormality Diagnosis          & Liver Lesion Diagnosis              & Liver      & Independent & 316   & 78  & 104  \\
MRNet             & Abnormality Diagnosis          & Knee Abnormality Diagnosis          & Knee       & Independent & 904   & 226 & 120  \\
Private-Knee\_ext & Abnormality Diagnosis          & Knee Abnormality Diagnosis          & Knee       & External    & 0     & 0   & 645  \\
Private-Knee\_int & Abnormality Diagnosis          & Knee Abnormality Diagnosis          & Knee       & Held-out    & 773   & 110 & 220  \\
\midrule
ADNI              & Disease Grading                & Alzheimer's Disease Grading         & Brain      & Held-out    & 1527  & 218 & 437  \\
CARE-Liver        & Disease Grading                & Liver Fibrosis Staging              & Liver      & Independent & 251   & 36  & 73   \\
OAI               & Disease Grading                & Knee Osteoarthritis Grading         & Knee       & Held-out    & 800   & 100 & 200  \\
OASIS 3           & Disease Grading                & Alzheimer's Disease Grading         & Brain      & Held-out    & 1583  & 226 & 453  \\
PI-CAI            & Disease Grading                & Prostate Cancer Grading             & Prostate   & Held-out    & 1000  & 150 & 350  \\
PPMI              & Disease Grading                & Parkinson's Disease Grading         & Brain      & Independent & 747   & 106 & 215  \\
\midrule
ADNI              & Progression Prediction         & AD Progression Prediction           & Brain      & Held-out    & 70    & 10  & 20   \\
OAI               & Progression Prediction         & OA Progression Prediction           & Knee       & Held-out    & 70    & 10  & 20   \\
\midrule
DUKE Liver        & Sequence Identification          & Sequence Identification               & Liver      & Independent & 73    & 10  & 22   \\
\midrule
IXI               & Cross-sequence Registration    & Brain Image Registration            & Brain      & Independent & 420   & 60  & 120  \\
OASIS             & Cross-sequence Registration    & Brain Image Registration            & Brain      & Held-out    & 416   & 0   & 19   \\
\midrule
ADNI              & Age Estimation                 & Age Estimation                      & Brain      & Held-out    & 1560  & 222 & 447  \\
OASIS 3           & Age Estimation                 & Age Estimation                      & Brain      & Held-out    & 1583  & 226 & 453  \\
openBHB           & Age Estimation                 & Age Estimation                      & Brain      & Independent & 2742  & 485 & 757  \\
\midrule
Private-Knee\_int & Radiological Report Generation & Report generation                   & Knee       & Held-out    & 140   & 20  & 40   \\
\midrule
ATLAS             & Lesion Segmentation            & Liver Cancer Segmentation           & Liver      & Independent & 42    & 6   & 12   \\
BraTS 2021        & Lesion Segmentation            & Brain Cancer Segmentation           & Brain      & Independent & 1000  & 0   & 251  \\
MAMA-MIA          & Lesion Segmentation            & Breast Tumor Segmentation           & Breast     & Independent & 1054  & 150 & 302  \\
PI-CAI            & Lesion Segmentation            & Prostate Cancer Segmentation        & Prostate   & Held-out    & 173   & 0   & 47   \\
Prostate158       & Lesion Segmentation            & Prostate Cancer Segmentation        & Prostate   & Independent & 120   & 19  & 19   \\
\midrule
ACDC              & Organ Segmentation             & Heart Structure Segmentation        & Heart      & Independent & 80    & 20  & 50   \\
AMOS              & Organ Segmentation             & Abdomen Multi-organ Segmentation    & Abdomen    & Independent & 70    & 10  & 20   \\
ATLAS             & Organ Segmentation             & Liver Segmentation                  & Liver      & External    & 0     & 0   & 60   \\
CHAOS             & Organ Segmentation             & Abdomen Multi-organ Segmentation    & Abdomen    & External    & 0     & 0   & 40   \\
DUKE Breast       & Organ Segmentation             & Breast Segmentation                 & Breast     & Independent & 70    & 10  & 20   \\
DUKE Liver        & Organ Segmentation             & Liver Segmentation                  & Liver      & Independent & 66    & 9   & 20   \\
EMIDEC            & Organ Segmentation             & Heart segmentation                  & Heart      & Independent & 70    & 10  & 20   \\
HaN-Seg           & Organ Segmentation             & Neck Segmentation                   & Neck       & Independent & 15    & 5   & 10   \\
MM-WHS            & Organ Segmentation             & Heart Structure segmentation        & Heart      & Independent & 14    & 2   & 4    \\
MSD-Cardiac       & Organ Segmentation             & Left Atrium Segmentation            & Heart      & External    & 0     & 0   & 20   \\
MSD-Hippocampus   & Organ Segmentation             & Hippocampus Segmentation            & Brain      & Independent & 213   & 50  & 131  \\
MSD-Prostate      & Organ Segmentation             & Prostate Segmentation               & Prostate   & Independent & 22    & 6   & 4    \\
OAI-ZIB           & Organ Segmentation             & Knee Structure Segmentation         & Knee       & Held-out    & 200   & 53  & 254  \\
PanSegData        & Organ Segmentation             & Pancreas Segmentation               & Abdomen    & External    & 0     & 0   & 767  \\
Private-Knee\_int & Organ Segmentation             & Cartilage Segmentation              & Knee       & Held-out    & 140   & 20  & 40   \\
PROMISE12         & Organ Segmentation             & Prostate Segmentation               & Prostate   & Independent & 50    & 30  & 20   \\
Prostate158       & Organ Segmentation             & Prostate Segmentation               & Prostate   & External    & 0     & 0   & 158  \\
SKM-TEA           & Organ Segmentation             & Knee Structure Segmentation         & Knee       & Independent & 86    & 33  & 36   \\
SPIDER            & Organ Segmentation             & Spine Segmentation                  & Spine      & Independent & 152   & 21  & 45   \\
Total Segmentator & Organ Segmentation             & Whole-body Multi-organ Segmentation & Whole-body & Independent & 430   & 61  & 125  \\
\bottomrule
\end{tabular}
\end{sidewaystable}

\subsection{Organ Segmentation}
We conducted comprehensive evaluations on organ segmentation across 20 tasks covering diverse anatomical regions, including the brain (e.g., MSD-Hippocampus), heart (e.g., ACDC, MSD-Cardiac), breast (e.g., DUKE Breast), abdomen (e.g., AMOS, CHAOS, PanSegData), prostate (e.g., MSD-Prostate, Prostate158, PROMISE12), knee (e.g., OAI-ZIB, Private-Knee), and spine (SPIDER). STable~\ref{tab:seg_organ} summarizes the Dice scores across these organ segmentation tasks.
STable~\ref{tab:seg_organ_ext} presents the segmentation performance on \textit{External} validation datasets to assess out-of-distribution generalization.
STable~\ref{tab:seg_organ_heart} reports class-wise segmentation results for the ACDC dataset, including the left ventricle, right ventricle, and myocardium.
STable~\ref{tab:seg_organ_knee} provides detailed results for individual knee structures in the OAI-ZIB dataset.
STable~\ref{tab:seg_organ_SPIDER} shows the segmentation performance of various spinal components in the SPIDER dataset.

\begin{sidewaystable}
\centering
\caption{The Dice score (\%) in organ segmentation tasks, including multiple organs across brain, breast, abdomen, and knee. The datasets are independent evaluations, except for the OAI-ZIB, which is a held-out set because its data is sourced from OAI~\cite{dsOAInevitt2006osteoarthritis}. Non-parametric bootstrapping with 1,000 bootstrap replicates is employed for statistical analysis, and we report the 95\% confidence intervals (CIs) in parentheses. For each dataset, the best-performing model is in bold, and the second-best-performing model is underlined. The $\Delta$ (Swin-B) metric quantifies the performance improvement of our pre-trained model over the Swin-B baseline trained from scratch. Significance marks (*, **, ***) indicate $p<0.05$, $p<0.01$, and $p<0.001$, respectively.}
\label{tab:seg_organ}
\begin{tabular}{llllll}
\toprule
Model\textbackslash   Dataset & ACDC                                          & AMOS                                          & DUKE Breast                                  & DUKE Liver                                   & EMIDEC                                       \\
nnUNet                        & \textbf{90.663 (86.751-94.575)}, p$<$0.001    & 86.186 (83.700-88.672), p$<$0.001             & 92.290 (90.077-94.503), p$<$0.001            & 71.349 (61.499-81.199), p$<$0.001            & 82.580 (80.282-84.878), p$<$0.01             \\
Swin-B                        & 86.545 (84.463-88.627), p$<$0.01              & 86.445 (83.819-89.071), p$<$0.001             & 93.105 (91.346-94.864), p$<$0.01             & 73.016 (67.132-78.900), p$<$0.001            & 81.814 (80.131-83.498), p$<$0.01             \\
MAE 3D                        & 86.201 (84.090-88.311), p$<$0.05              & 86.625 (83.693-89.557), p$<$0.001             & 93.891 (90.891-96.892), p$<$0.001            & 72.917 (66.044-79.790), p$<$0.05             & 83.735 (80.815-86.655), p$<$0.001            \\
MoCo V3                       & 86.894 (85.174-88.614), p$<$0.001             & \underline{87.523 (86.261-88.786)}, p$<$0.05  & 92.436 (90.736-94.135), p$<$0.01             & \underline{73.505 (65.488-81.522)}, p$<$0.01 & 80.667 (77.838-83.495), p$<$0.001            \\
SwinUNETR                     & 87.425 (85.939-88.910), p$<$0.01              & 87.137 (84.385-89.889), p$<$0.001             & \textbf{94.069 (90.659-97.478)}, p$<$0.05    & 71.913 (65.733-78.092), p$<$0.001            & 83.449 (81.371-85.526), p$<$0.001            \\
BrainSegF.                    & 86.797 (83.855-89.738), p$<$0.01              & 87.200 (84.562-89.838), p$<$0.01              & \underline{93.947 (91.050-96.843)}, p$<$0.01 & 71.444 (64.764-78.123), p$<$0.001            & \underline{84.178 (82.109-86.247)}, p$<$0.01 \\
Ours                          & \underline{87.715 (85.488-89.941)}            & \textbf{87.633 (84.807-90.459)}               & 93.310 (91.010-95.610)                       & \textbf{75.703 (65.580-85.827)}              & \textbf{85.094 (83.065-87.123)}              \\
\midrule
$\Delta$ (Swin-B)             & 1.170 (**)                                    & 1.188 (***)                                   & 0.205 (**)                                   & 2.687 (***)                                  & 3.279 (**)                                   \\
\bottomrule
\addlinespace[1pt]
\toprule
Model\textbackslash Dataset   & HaN-Seg                                       & MM-WHS                                        & MSD-Hippocampus                              & MSD-Prostate                                 & OAI-ZIB                                      \\
\midrule
nnUNet                        & 75.258 (72.097-78.420), p$<$0.05              & 72.870 (65.733-80.008), p$<$0.01              & 83.591 (81.543-85.639), p$<$0.001            & 74.268 (72.504-76.032), p$<$0.001            & 92.765 (90.531-94.999), p$<$0.001            \\
Swin-B                        & 75.052 (72.383-77.720), p$<$0.001             & 69.794 (64.454-75.135), p$<$0.001             & 83.856 (80.929-86.782), p$<$0.001            & 74.645 (72.133-77.157), p$<$0.001            & 92.403 (90.493-94.313), p$<$0.01             \\
MAE 3D                        & 75.381 (73.092-77.670), p$<$0.01              & 71.606 (66.371-76.841), p$<$0.001             & 84.056 (80.720-87.391), p$<$0.01             & 75.757 (73.507-78.006), p$<$0.05             & 92.711 (90.685-94.737), p$<$0.001            \\
MoCo V3                       & 76.072 (73.258-78.887), p$<$0.01              & 69.078 (61.379-76.776), p$<$0.05              & \underline{84.431 (81.117-87.746)}, p$<$0.01 & 73.289 (72.085-74.493), p=0.058              & 92.925 (91.657-94.193), p$<$0.01             \\
SwinUNETR                     & \underline{76.102 (73.554-78.651)}, p$<$0.01  & \underline{72.892 (67.404-78.381)}, p$<$0.001 & 84.205 (80.688-87.721), p$<$0.01             & \textbf{78.087 (75.759-80.415)}, p$<$0.01    & 93.125 (91.796-94.454), p$<$0.01             \\
BrainSegF.                    & 75.806 (72.014-79.599), p$<$0.01              & 70.873 (64.765-76.981), p$<$0.05              & 83.603 (80.674-86.531), p$<$0.001            & 76.248 (74.025-78.472), p$<$0.01             & \underline{93.168 (90.350-95.986)}, p$<$0.05 \\
Ours                          & \textbf{77.176 (75.125-79.227)}               & \textbf{73.115 (68.088-78.143)}               & \textbf{84.570 (83.420-85.720)}              & \underline{77.580 (75.708-79.452)}           & \textbf{93.203 (91.186-95.219)}              \\
\midrule
$\Delta$ (Swin-B)             & 2.124 (***)                                   & 3.321 (***)                                   & 0.714 (***)                                  & 2.935 (***)                                  & 0.799 (**)                                   \\
\bottomrule
\addlinespace[1pt]
\toprule
Model\textbackslash Dataset   & Private-Knee\_int                              & PROMISE12                                     & SKM-TEA                                      & SPIDER                                       & Total Segmentator                            \\
\midrule
nnUNet                        & 86.940 (82.253-91.627), p$<$0.001             & 83.158 (80.974-85.341), p$<$0.001             & \underline{87.581 (84.753-90.408)}, p$<$0.01 & 90.235 (88.968-91.502), p$<$0.001            & 74.090 (70.879-77.301), p=0.071              \\
Swin-B                        & 86.190 (83.344-89.036), p$<$0.001             & 83.406 (80.244-86.569), p$<$0.01              & 85.805 (82.576-89.034), p$<$0.001            & 89.717 (87.765-91.670), p$<$0.001            & 74.806 (71.992-77.621), p$<$0.001            \\
MAE 3D                        & 85.808 (82.952-88.665), p$<$0.01              & 83.629 (78.966-88.293), p$<$0.001             & 85.755 (81.676-89.835), p$<$0.01             & 89.477 (88.257-90.697), p$<$0.01             & 75.670 (70.976-80.365), p$<$0.01             \\
MoCo V3                       & 86.900 (81.234-92.566), p$<$0.001             & 84.624 (81.624-87.625), p$<$0.001             & 86.257 (83.290-89.225), p$<$0.01             & 88.885 (86.370-91.400), p$<$0.001            & 73.934 (72.366-75.503), p$<$0.001            \\
SwinUNETR                     & \underline{87.142 (80.476-93.808)}, p$<$0.001 & 84.199 (81.824-86.574), p$<$0.001             & 87.307 (82.781-91.834), p$<$0.001            & 90.438 (88.610-92.267), p$<$0.01             & \textbf{78.011 (76.827-79.195)}, p$<$0.05    \\
BrainSegF.                    & 86.815 (82.164-91.467), p$<$0.001             & \underline{85.085 (81.581-88.589)}, p$<$0.001 & 85.652 (83.186-88.118), p$<$0.05             & \underline{90.493 (88.624-92.363)}, p$<$0.05 & 76.625 (73.061-80.189), p$<$0.001            \\
Ours                          & \textbf{87.300 (83.254-91.346)}               & \textbf{85.180 (81.389-88.970)}               & \textbf{88.993 (85.203-92.784)}              & \textbf{90.906 (89.850-91.962)}              & \underline{77.379 (74.647-80.111)}           \\
\midrule
$\Delta$ (Swin-B)             & 1.110 (***)                                   & 1.773 (**)                                    & 3.188 (***)                                  & 1.189 (***)                                  & 2.572 (***)                                 
\end{tabular}
\end{sidewaystable}

% The OAI-ZIB contains segmentation tasks of the femoral bone (FB), tibial bone (TB), and the respective femoral and tibial cartilage (FC, TC).

\begin{table*}[ht]
\centering
\caption{The Dice score (\%) on external segmentation test set. The $\Delta$ (Swin-B) metric quantifies the performance improvement of our pre-trained model over the Swin-B baseline trained from scratch. Significance marks (*, **, ***) indicate $p<0.05$, $p<0.01$, and $p<0.001$, respectively.}
\label{tab:seg_organ_ext}

    \begin{tabular}{llllll}
    \toprule
    Model\textbackslash   Dataset           & CHAOS         & MSD-Cardiac     & Prostate 158  & ATLAS & PanSegData \\
    \midrule
    nnUNet           & 89.59 (±0.86) & 88.55 (±0.18) & 78.53 (±3.04) & 68.52 (±0.91)     & 78.84 (±2.41)      \\
    Swin-B           & 89.19 (±1.24) & 88.67 (±0.20) & 78.73 (±1.88) & 68.37 (±2.72)     & 76.69 (±1.42)      \\
    MAE 3D           & 88.82 (±1.55) & 88.14 (±0.47) & 78.69 (±2.02) & 69.17 (±1.36)     & 77.01 (±2.27)      \\
    MoCo V3          & 88.44 (±1.57) & 88.48 (±0.38) & 79.66 (±1.90) & 68.16 (±0.80)     & 77.75 (±1.10)      \\
    SwinUNETR        & 90.11 (±1.61) & 89.07 (±0.58) & 79.17 (±1.60) & 68.30 (±2.99)     & 78.97 (±3.51)      \\
    BrainSegF.       & 89.84 (±1.71) & 88.90 (±0.52) & 78.84 (±3.49) & 69.54 (±2.89)     & 79.19 (±3.65)      \\
    Ours             & \textbf{90.68 (±1.53)} & \textbf{89.76 (±0.28)} & \textbf{80.17 (±2.22)} & \textbf{70.22 (±1.49)}     & \textbf{79.21 (±2.21)}      \\
    \midrule
    $\Delta$(Swin-B) & 1.48 (*)      & 1.09 (***)    & 1.44 (**)     & 1.85 (***)                               & 2.52 (***)                         \\
    \bottomrule                      
    \end{tabular}
\end{table*}

%******************detail
\begin{table*}[ht]
\centering
\caption{Detailed heart organ segmentation result on the ACDC dataset. Reported by Dice score (\%) for each class. The $\Delta$ (Swin-B) metric quantifies the performance improvement of our pre-trained model over the Swin-B baseline trained from scratch. Significance marks (*, **, ***) indicate $p<0.05$, $p<0.01$, and $p<0.001$, respectively.}
\label{tab:seg_organ_heart}
\begin{tabular}{llll}
\toprule
Model\textbackslash   Dataset & ACDC Left Ventricle    & ACDC Right Ventricle   & ACDC Myocardium        \\ \midrule
nnUNet                        & \textbf{91.84 (±2.73)} & \textbf{88.68 (±2.08)} & \textbf{91.47 (±1.97)} \\
Swin-B                        & 87.78 (±2.00)          & 83.71 (±2.14)          & 87.15 (±2.44)          \\
MAE 3D                        & 87.72 (±1.66)          & 84.44 (±3.38)          & 86.44 (±1.48)          \\
MoCo V3                       & 88.20 (±1.71)          & 84.92 (±2.04)          & 87.55 (±1.94)          \\
SwinUNETR                     & 88.84 (±1.84)          & 85.37 (±2.09)          & 88.07 (±2.53)          \\
BrainSegF.                    & 87.98 (±2.88)          & 84.96 (±2.05)          & 87.45 (±2.83)          \\
Ours                          & 89.19 (±1.47)          & 85.69 (±1.15)          & 88.26 (±2.20)          \\
\midrule
$\Delta$(Swin-B)              & 1.41 (**)              & 0.98 (**)              & 1.11 (**)              \\ \bottomrule
\end{tabular}
\end{table*}

\begin{table*}[ht]
\centering
\caption{Detailed segmentation results on the OAI-ZIB dataset for femoral cartilage (FC), tibial cartilage (TC), femoral bone (FB), and tibial bone (TB), reported by Dice score (\%). The $\Delta$ (Swin-B) metric quantifies the performance improvement of our pre-trained model over the Swin-B baseline trained from scratch. Significance marks (*, **, ***) indicate $p<0.05$, $p<0.01$, and $p<0.001$, respectively.}
\label{tab:seg_organ_knee}
\begin{tabular}{lllll}
\toprule
Model\textbackslash   Dataset & OAI-ZIB FC             & OAI-ZIB TC             & OAI-ZIB FB             & OAI-ZIB TB             \\
\midrule
nnUNet                        & 89.46 (±3.22)          & 85.16 (±1.82)          & 98.31 (±2.67)          & 98.13 (±1.71)          \\
Swin-B                        & 89.30 (±2.96)          & 83.19 (±3.19)          & 98.31 (±2.22)          & 98.81 (±3.76)          \\
MAE 3D                        & 89.62 (±3.36)          & 84.66 (±2.55)          & 98.00 (±2.16)          & 98.57 (±2.49)          \\
MoCo V3                       & 89.79 (±1.98)          & 85.63 (±1.07)          & 98.32 (±1.67)          & 97.95 (±1.95)          \\
SwinUNETR                     & 89.71 (±1.47)          & 85.33 (±1.53)          & \textbf{98.59 (±2.89)} & 98.87 (±1.78)          \\
BrainSegF.                    & 89.74 (±2.93)          & 85.60 (±3.40)          & 98.49 (±1.02)          & 98.84 (±1.57)          \\
Ours                          & \textbf{89.93 (±3.07)} & \textbf{85.64 (±2.26)} & 98.37 (±2.42)          & \textbf{98.87 (±3.16)} \\
\midrule
$\Delta$(Swin-B)              & 0.63 (*)               & 2.45 (*)               & 0.06 (*)               & 0.06 (**)             \\
\bottomrule
\end{tabular}
\end{table*}

\begin{table*}[ht]
\centering
\caption{Detailed spine segmentation result on SPIDER dataset. Reported by Dice score (\%) for each class. The $\Delta$ (Swin-B) metric quantifies the performance improvement of our pre-trained model over the Swin-B baseline trained from scratch. Significance marks (*, **, ***) indicate $p<0.05$, $p<0.01$, and $p<0.001$, respectively.}
\label{tab:seg_organ_SPIDER}
\begin{tabular}{llll}
\toprule
Model\textbackslash   Dataset & SPIDER-Vertebrae       & SPIDER-IVDs            & SPIDER-Spinal Canal    \\
\midrule
nnUNet                        & 93.11 (±1.87)          & 84.50 (±1.89)          & 93.10 (±1.40)          \\
Swin-B                        & 92.86 (±3.56)          & 84.14 (±1.27)          & 92.15 (±1.13)          \\
MAE 3D                        & 92.53 (±3.51)          & 83.38 (±1.92)          & 92.52 (±1.33)          \\
MoCo V3                       & 92.31 (±1.70)          & 83.30 (±1.48)          & 91.04 (±3.07)          \\
SwinUNETR                     & \textbf{93.41 (±2.40)} & 85.09 (±1.49)          & 92.82 (±1.05)          \\
BrainSegF.                    & 93.08 (±1.80)          & 85.08 (±3.17)          & 93.32 (±2.17)          \\
Ours                          & 93.37 (±2.93)          & \textbf{85.38 (±1.69)} & \textbf{93.97 (±2.54)} \\
\midrule
$\Delta$(Swin-B)              & 0.51 (**)              & 1.23 (*)               & 1.82 (**)             \\
\bottomrule
\end{tabular}
\end{table*}

\clearpage

\subsection{Lesion Segmentation}
We conducted comprehensive evaluations on lesion segmentation across five tasks, encompassing diverse disease types and anatomical sites, including liver cancer (ATLAS), brain metastasis (BraTS), breast cancer (MAMA-MIA), and prostate cancer (PI-CAI and Prostate158).
STable~\ref{tab:seg_tumor} summarizes the Dice scores across these lesion segmentation tasks, highlighting the performance of our model compared to strong baselines.
STable~\ref{tab:seg_lesion_detailed} presents class-wise segmentation results for multi-class lesion tasks, including tumor subregions in the BraTS dataset (enhancing tumor, tumor core, whole tumor) and breast tissue structures in the DUKE Breast dataset (fibroglandular tissue and vessels).
\begin{table*}[ht]
\centering
\caption{The Dice score (\%) in 5 lesion segmentation tasks, including liver cancer (ATLAS), brain metastasis (BraTS), breast cancer (MAMA-MIA), prostate cancer (PI-CAI and Prostate158). Non-parametric bootstrapping with 1,000 bootstrap replicates is employed for statistical analysis, and we report the 95\% confidence intervals (CIs) in parentheses. For each dataset, the best-performing model is in bold, and the second-best-performing model is underlined. The $\Delta$ (Swin-B) metric quantifies the performance improvement of our pre-trained model over the Swin-B baseline trained from scratch. Significance marks (*, **, ***) indicate $p<0.05$, $p<0.01$, and $p<0.001$, respectively.}
\label{tab:seg_tumor}
\begin{tabular}{lll}
\toprule
Model\textbackslash   Dataset & ATLAS                                        & BraTS 2021                                   \\ \midrule
nnUNet                        & 74.300 (71.262-77.337), p$<$0.01             & 88.698 (81.309-96.088), p$<$0.01             \\
Swin-B                        & 75.791 (72.975-78.607), p$<$0.01             & 88.795 (81.813-95.778), p$<$0.01             \\
MAE 3D                        & 74.658 (71.317-77.998), p$<$0.05             & 88.514 (80.439-96.589), p$<$0.001            \\
MoCo V3                       & 73.449 (69.497-77.401), p$<$0.001            & 89.355 (80.433-98.277), p$<$0.01             \\
SwinUNETR                     & \underline{76.595 (72.921-80.268)}, p$<$0.01 & 89.510 (81.481-97.540), p$<$0.01             \\
BrainSegF.                    & 76.382 (71.753-81.010), p$<$0.01             & \underline{89.832 (80.412-99.251)}, p$<$0.05 \\
Ours                          & \textbf{77.379 (73.627-81.130)}              & \textbf{90.583 (82.361-98.805)}              \\
\midrule
$\Delta$ (Swin-B)             & 1.588 (**)                                   & 1.788 (**)                                   \\
\bottomrule
\addlinespace[1pt]
\toprule
Model\textbackslash Dataset   & MAMA-MIA                                     & PI-CAI                                       \\
nnUNet                        & 75.824 (69.188-82.460), p$<$0.001            & 73.673 (63.010-84.336), p$<$0.01             \\
Swin-B                        & 74.302 (68.369-80.234), p$<$0.01             & 75.892 (67.648-84.136), p$<$0.05             \\
MAE 3D                        & 75.488 (67.462-83.514), p=0.096              & 74.679 (63.481-85.877), p$<$0.01             \\
MoCo V3                       & 73.162 (66.005-80.319), p$<$0.05             & 74.417 (65.198-83.636), p$<$0.001            \\
SwinUNETR                     & \underline{76.585 (68.961-84.208)}, p$<$0.05 & \underline{76.934 (66.372-87.495)}, p$<$0.05 \\
BrainSegF.                    & 74.224 (67.972-80.477), p$<$0.01             & 76.614 (65.383-87.846), p$<$0.01             \\
Ours                          & \textbf{77.562 (71.326-83.798)}              & \textbf{77.140 (66.457-87.823)}              \\
\midrule
$\Delta$ (Swin-B)             & 3.260 (**)                                   & 1.248 (*)                                    \\
\bottomrule
\addlinespace[1pt]
\toprule
Model\textbackslash Dataset   & Prostate158                                  &                                              \\
nnUNet                        & \multicolumn{2}{l}{87.320 (80.551-94.089),   p$<$0.05}                                      \\
Swin-B                        & \multicolumn{2}{l}{86.698 (77.583-95.813),   p$<$0.01}                                      \\
MAE 3D                        & \multicolumn{2}{l}{88.658 (80.188-97.129),   p$<$0.001}                                     \\
MoCo V3                       & \multicolumn{2}{l}{87.587 (81.580-93.595),   p$<$0.05}                                      \\
SwinUNETR                     & \multicolumn{2}{l}{87.930 (80.079-95.781),   p$<$0.001}                                     \\
BrainSegF.                    & \multicolumn{2}{l}{\underline{88.740   (81.903-95.577)}, p$<$0.001}                         \\
Ours                          & \multicolumn{2}{l}{\textbf{88.760   (80.357-97.163)}}                                       \\
\midrule
$\Delta$ (Swin-B)             & \multicolumn{2}{l}{2.062 (**)}                                                              \\ \bottomrule
\bottomrule
\end{tabular}
\end{table*}

\begin{table*}[ht]
\centering
\caption{Detailed lesion segmentation result, including multi-class segmentation on brain metastasis (BraTS) and breast cancer (DUKE Breast). For the BraTS dataset, the segmented targets include tumor core (TC), whole tumor (WT), and enhancing tumor (ET). Reported by Dice score (\%) for each class. The $\Delta$ (Swin-B) metric quantifies the performance improvement of our pre-trained model over the Swin-B baseline trained from scratch. Significance marks (*, **, ***) indicate $p<0.05$, $p<0.01$, and $p<0.001$, respectively.}
\label{tab:seg_lesion_detailed}
\resizebox{\textwidth}{!}{
\begin{tabular}{llllll}
\toprule
Model\textbackslash   Dataset           & BraTS ET      & BraTS TC      & BraTS WT      & DUKE Breast FGT & DUKE Breast Vessle \\
\midrule
nnUNet           & 84.97 (±2.68) & 89.86 (±2.66) & 91.26 (±1.29) & 84.54 (±2.27)   & 65.66 (±2.93)      \\
Swin-B           & 84.14 (±1.76) & 89.51 (±2.42) & 92.74 (±1.61) & 85.87 (±3.05)   & 64.15 (±2.96)      \\
MAE 3D           & 85.16 (±1.57) & 88.86 (±1.55) & 91.52 (±1.01) & 85.80 (±2.68)   & 64.25 (±4.27)      \\
MoCo V3          & 83.91 (±1.84) & 91.54 (±2.19) & 92.61 (±2.21) & 84.50 (±3.77)   & 67.50 (±3.35)      \\
SwinUNETR        & 84.51 (±2.57) & 90.57 (±2.14) & 93.45 (±1.35) & 87.09 (±1.84)   & 64.79 (±2.73)      \\
BrainSegF.       & 85.27 (±1.95) & 90.95 (±2.19) & 93.28 (±1.74) & 86.06 (±2.65)   & 65.85 (±1.54)      \\
Ours & \textbf{86.44 (±3.37)} & \textbf{91.83 (±1.62)} & \textbf{93.48 (±2.15)} & \textbf{87.38 (±2.07)} & \textbf{67.92 (±3.45)} \\
\midrule
$\Delta$(Swin-B) & 2.30 (**)     & 2.32 (**)     & 0.74 (**)     & 1.51 (***)      & 3.77 (**)         \\
\bottomrule
\end{tabular}
}
\end{table*}

\clearpage
\subsection{Abnormality Diagnosis}

The abnormality diagnosis is conducted for knee abnormality classification and liver cancer classification. The knee abnormality classification tasks on the Private-Knee dataset and MRNet all involve multiple binary classification problems, whereas the liver cancer classification on the LLD-MMRI dataset constitutes a multi-class classification task.

\noindent\textbf{Private-Knee} The private knee dataset contains 12 different types of knee abnormalities and includes T1W, T2W and PDW sequences. We follow our previous work~\cite{qiu2024learning} in separating the train, validation, and testing splits, as well as the external dataset. The data in this study covers 12 types of knee abnormalities, including meniscal tear (MENI); anterior cruciate ligament tear (ACL); cartilage damage (CART); posterior cruciate ligament injury (PCL); medial collateral ligament injury (MCL); lateral collateral ligament injury (LCL); joint effusion (EFFU); bone contusion (CONT); synovial plica (PLICA); cyst (CYST); infrapatellar fat pad injury (IFP); and patellar retinaculum injury (PR). The Private-Knee\_ext dataset encompasses four centers. We deploy the model trained on Private-Knee\_int for external validation.

\noindent\textbf{MRNet} The MRNet~\cite{dsMRNetbien2018deep} dataset encompasses three classification tasks: ACL tears (ACL), meniscal tears (MENI), and other abnormalities. The data includes MRI sequences of coronal T1W, coronal T2W with fat saturation, sagittal PDW, sagittal T2W with fat saturation, and axial PDW with fat saturation. We employed the provided validation set as the testing split and partitioned the original training set into new training and validation subsets at an 8:2 ratio.

\noindent\textbf{LLD-MMRI} The LLD-MMRI~\cite{dsLLDlou2025sdr} dataset is designed for the diagnosis of liver abnormalities, encompassing hepatic hemangioma, intrahepatic cholangiocarcinoma, hepatic abscess, hepatic metastasis, hepatic cyst, focal nodular hyperplasia, and hepatocellular carcinoma. The data includes multi-phasic DCE (Dynamic Contrast-Enhanced) and non-contrast T2W sequences. We retained the provided data split for training, validation, and testing.

\noindent The corresponding results are shown in STable~\ref{tab:abn_dia}.

\begin{table*}[ht]
\centering
\caption{ The classification accuracy (ACC, \%) in abnormality diagnosis for liver (LLD-MMRI dataset) and knee (MRNet and Private-Knee datasets) is presented. Non-parametric bootstrapping with 1,000 bootstrap replicates is employed for statistical analysis, and we report the 95\% confidence intervals (CIs) in parentheses. For each dataset, the best-performing model is in bold, and the second-best-performing model is underlined. The $\Delta$ (Swin-B) metric quantifies the performance improvement of our pre-trained model over the Swin-B baseline trained from scratch. Significance marks (*, **, ***) indicate $p<0.05$, $p<0.01$, and $p<0.001$, respectively.}
\label{tab:abn_dia}
\centering
\begin{tabular}{lllll}
\toprule
Model\textbackslash   Dataset & LLD-MMRI                                      & MRNet                                         \\ \midrule
ResNet3D                      & 68.365 (60.614-76.117), p$<$0.001             & 78.959 (74.620-83.297), p$<$0.001             \\
Swin-B                        & \underline{73.712 (67.365-80.058)}, p$<$0.001 & 77.781 (75.373-80.190), p$<$0.001             \\
MAE 3D                        & 70.955 (63.776-78.134), p$<$0.001             & 77.055 (73.794-80.316), p$<$0.05              \\
MoCo V3                       & 68.269 (59.809-76.730), p$<$0.001             & 77.142 (74.702-79.582), p$<$0.001             \\
SwinUNETR                     & 72.877 (65.574-80.180), p=0.623               & \underline{79.510 (77.536-81.484)}, p$<$0.001 \\
BrainSegF.                    & 73.137 (65.559-80.714), p$<$0.001             & 76.764 (74.540-78.988), p$<$0.001             \\
Ours                          & \textbf{75.836 (68.668-83.003)}               & \textbf{80.047 (75.759-84.335)}               \\
\midrule
$\Delta$ (Swin-B)             & 2.124 (***)                                   & 2.266 (***)                                   \\
\bottomrule
\addlinespace[1pt]
\toprule
Model\textbackslash Dataset   & Private-Knee\_int                              & Private-Knee\_ext                              \\
\midrule
ResNet3D                      & 75.399 (70.866-79.933), p$<$0.001             & 69.040 (61.675-76.404), p$<$0.001             \\
Swin-B                        & 76.942 (75.502-78.383), p$<$0.001             & 69.540 (61.922-77.158), p$<$0.01              \\
MAE 3D                        & 75.798 (69.243-82.352), p$<$0.05              & 69.852 (63.698-76.006), p$<$0.01              \\
MoCo V3                       & 78.071 (72.659-83.483), p=0.140               & 70.806 (63.305-78.307), p=0.123               \\
SwinUNETR                     & 78.263 (75.912-80.614), p=0.224               & 70.233 (65.297-75.169), p$<$0.05              \\
BrainSegF.                    & \underline{79.478 (74.197-84.759)}, p=0.342   & \underline{70.883 (64.050-77.715)}, p=0.137   \\
Ours                          & \textbf{80.879 (75.723-86.034)}               & \textbf{71.816 (65.517-78.114)}               \\
\midrule
$\Delta$ (Swin-B)             & 3.936 (***)                                   & 2.276 (**)                                    \\ \bottomrule
\end{tabular}
\end{table*}

\clearpage

\subsection{Disease Grading}

The automatic disease grading aims to assess the severity or stage of diseases. We conducted comprehensive evaluations on multiple diseases and lesions, including Alzheimer's disease (AD) using the ADNI and OASIS datasets, Parkinson's disease (PD) using the PPMI dataset, osteoarthritis (OA) using the OAI dataset, and prostate cancer (PCa) using the PI-CAI dataset.

\noindent\textbf{PPMI} The PPMI dataset~\cite{dsPPMImarek2018parkinson} encompasses multiple stages of subjects across different phases of Parkinson's progression. Specifically, we selected subjects with 3D T1-weighted sequences. The resulting subject groups include 'Prodromal', 'PD', and 'Control'. We divided the selected data into training, validation, and testing sets at a ratio of 7:1:2.

\noindent\textbf{PI-CAI} The PI-CAI dataset~\cite{dsPICAIsaha2024artificial} comprises 1,500 clinical cases, each featuring multi-parametric MRI sequences including T2W, DWI, and ADC maps. Leveraging the provided masks available in the dataset, we concentrate our analysis on cropped tumor patches to perform a classification task aimed at distinguishing clinically significant prostate cancer (csPCa).

\noindent\textbf{ADNI} For the ADNI dataset, we conducted classification among cognitively normal (CN), mild cognitive impairment (MCI), and Alzheimer’s disease (AD) groups. Following the approach in~\cite{majee2024enhancing}, we utilize the "ADNI1\_Complete 3Yr 1.5T" subset, which encompasses clinical visits of patients screened over a 6-month to 3-year period, comprising 2182 cases.

\noindent\textbf{OASIS 3} Evaluation on the OASIS 3 dataset is conducted in a single-sequence approach to classify three distinct AD stages. The stages are categorized based on the total Clinical Dementia Rating (CDR) score~\cite{dsOASISlamontagne2019oasis}, where each image is mapped to the nearest corresponding clinical timepoint to determine its score: CN (CDR=0), MCI (CDR=0.5), and AD (CDR$\geq$1).
The dataset is partitioned according to different MRI scanning sessions, with the resulting split ratios for the training, validation, and testing sets being 7:1:2, respectively.

\noindent\textbf{OAI} We follow the previous work on osteoarthritis grading~\cite{guida2021knee} and evaluate the grading performance for KL=0 to KL=4. We selected subjects from the OAI dataset using the provided subject ID list. To maintain the data split ratio, we swapped the validation and testing datasets of the reference paper, resulting in a testing dataset of 200 subjects.

\noindent The corresponding results are reported in STable~\ref{tab:grading_suppl}.

% \subsection{Progression prediction}

 % \textbf{PI-CAI} The segmentation includes 1295 cases. We perform the Clinically significant prostate cancer (csPCa)  segmentation evaluation on T2-weghted used the annotations that is resampled to axial T2-weighted scan . 

\begin{table*}[ht]
\centering
\caption{The classification accuracy (ACC, \%) for disease grading, including Alzheimer's Disease (ADNI, OASIS 3), Liver Fibrosis Staging (CARE-Liver), Knee Osteoarthritis (OAI), Prostate Cancer (PI-CAI) and Parkinson's Disease (PPMI). Non-parametric bootstrapping with 1,000 bootstrap replicates is employed for statistical analysis, and we report the 95\% confidence intervals (CIs) in parentheses. For each dataset, the best-performing model is in bold, and the second-best-performing model is underlined. The $\Delta$ (Swin-B) metric quantifies the performance improvement of our pre-trained model over the Swin-B baseline trained from scratch. Significance marks (*, **, ***) indicate $p<0.05$, $p<0.01$, and $p<0.001$, respectively.}
\label{tab:grading_suppl}
\begin{tabular}{lll}
\toprule
Model\textbackslash   Dataset & ADNI                                          & CARE-Liver                                   \\ \midrule
ResNet3D                      & 87.730 (83.866-91.594), p$<$0.01              & 74.198 (69.926-78.470), p$<$0.001            \\
Swin-B                        & 86.368 (82.813-89.924), p$<$0.05              & 78.106 (73.189-83.023), p$<$0.001            \\
MAE 3D                        & 88.407 (85.917-90.896), p=0.070               & 78.012 (73.099-82.926), p$<$0.01             \\
MoCo V3                       & 86.041 (83.639-88.442), p$<$0.001             & 76.537 (71.681-81.394), p$<$0.01             \\
SwinUNETR                     & 87.400 (85.622-89.178), p$<$0.001             & \underline{78.183 (74.567-81.799)}, p$<$0.01 \\
BrainSegF.                    & \textbf{89.032 (87.164-90.900)}, p$<$0.05     & 77.242 (71.511-82.973), p$<$0.05             \\
Ours                          & \underline{88.423 (86.257-90.588)}            & \textbf{79.863 (75.484-84.242)}              \\
\midrule
$\Delta$ (Swin-B)             & 2.054 (*)                                     & 1.757 (***)                                  \\
\bottomrule
\addlinespace[1pt]
\toprule
Model\textbackslash Dataset   & OAI                                           & OASIS 3                                      \\
ResNet3D                      & 62.039 (59.838-64.240), p$<$0.001             & 83.036 (79.163-86.909), p$<$0.001             \\
Swin-B                        & 63.201 (58.725-67.677), p$<$0.01              & 82.744 (79.386-86.103), p$<$0.001             \\
MAE 3D                        & \underline{64.079 (61.069-67.089)}, p$<$0.05  & \underline{83.982 (81.714-86.249)}, p$<$0.001 \\
MoCo V3                       & 63.179 (60.588-65.770), p$<$0.01              & 81.311 (77.325-85.298), p$<$0.001             \\
SwinUNETR                     & 62.050 (58.474-65.626), p$<$0.001             & 82.817 (79.361-86.273), p$<$0.001             \\
BrainSegF.                    & 62.290 (58.262-66.318), p$<$0.01              & 83.536 (80.158-86.914), p$<$0.001             \\
Ours                          & \textbf{64.202 (61.299-67.104)}               & \textbf{85.052 (82.577-87.527)}               \\
\midrule
$\Delta$ (Swin-B)             & 1.000 (**)                                    & 2.308 (***)                                   \\ 
\bottomrule
\addlinespace[1pt]
\toprule
Model\textbackslash Dataset   & PI-CAI                                        & PPMI                                         \\
ResNet3D                      & 85.229 (80.370-90.087), p$<$0.01              & 69.384 (64.945-73.823), p$<$0.001            \\
Swin-B                        & 84.940 (83.830-86.050), p$<$0.01              & 69.127 (64.858-73.396), p$<$0.05             \\
MAE 3D                        & 85.145 (81.165-89.125), p$<$0.01              & 69.966 (66.132-73.801), p$<$0.05             \\
MoCo V3                       & 85.805 (80.379-91.231), p$<$0.001             & 69.948 (66.574-73.323), p=0.161              \\
SwinUNETR                     & \underline{86.749 (80.741-92.757)}, p$<$0.001 & \underline{71.001 (67.780-74.222)}, p$<$0.05 \\
BrainSegF.                    & 86.210 (79.053-93.367), p$<$0.01              & 69.904 (67.678-72.131), p$<$0.001            \\
Ours                          & \textbf{86.782 (79.397-94.167)}               & \textbf{71.753 (69.644-73.861)}              \\
\midrule
$\Delta$ (Swin-B)             & 1.842 (**)                                    & 2.626 (*)                                    \\ \bottomrule
\end{tabular}
\end{table*}
\clearpage

\subsection{Progression Prediction}
We conducted comprehensive evaluations on disease progression prediction across two chronic disease cohorts, including Alzheimer's disease (ADNI) and knee osteoarthritis (OAI). These tasks aim to forecast the future trajectory of diseases based on baseline MRI, supporting early risk stratification and longitudinal management.
STable~\ref{tab:progression_suppl} summarizes the classification accuracy across both datasets, demonstrating that \modelname~consistently outperforms strong baselines. Notably, our model achieves the best performance on both ADNI and OAI cohorts, with statistically significant improvements in the ADNI setting.

\begin{table*}[ht]
\centering
\caption{The classification accuracy (ACC, \%) for progression prediction on Alzheimer's disease (ADNI) and knee osteoarthritis (OAI). Non-parametric bootstrapping with 1,000 bootstrap replicates is employed for statistical analysis, and we report the 95\% confidence intervals (CIs) in parentheses. For each dataset, the best-performing model is in bold, and the second-best-performing model is underlined. The $\Delta$ (Swin-B) metric quantifies the performance improvement of our pre-trained model over the Swin-B baseline trained from scratch. Significance marks (*, **, ***) indicate $p<0.05$, $p<0.01$, and $p<0.001$, respectively.}
\label{tab:progression_suppl}
\begin{tabular}{lll}
\toprule
Model\textbackslash   Dataset & ADNI                                         & OAI                                          \\ \midrule
ResNet3D                      & 73.257 (69.011-77.503), p$<$0.001            & 58.859 (52.821-64.896), p$<$0.001            \\
Swin-B                        & 75.801 (71.790-79.812), p$<$0.001            & 61.866 (53.029-70.703), p=0.064              \\
MAE 3D                        & 75.662 (71.635-79.689), p$<$0.001            & \underline{62.267 (56.474-68.061)}, p$<$0.05 \\
MoCo V3                       & 74.526 (69.569-79.484), p$<$0.01             & 61.625 (52.260-70.990), p$<$0.05             \\
SwinUNETR                     & 76.065 (71.326-80.805), p$<$0.001            & 62.247 (55.178-69.315), p$<$0.001            \\
BrainSegF.                    & \underline{76.753 (72.806-80.700)}, p$<$0.01 & 61.824 (52.415-71.233), p$<$0.001            \\
Ours                          & \textbf{77.338 (72.638-82.038)}              & \textbf{63.615 (55.002-72.227)}              \\ \midrule
$\Delta$ (Swin-B)             & 1.537 (***)                                  & 1.749                                        \\ \bottomrule
\end{tabular}
\end{table*}

\clearpage

\subsection{Sequence Identification}
We evaluated \modelname~on the sequence identification task, which aims to classify the specific MRI sequence type from a given scan—a critical step for automated quality control, data curation, and downstream task adaptation.
STable~\ref{tab:sequence_identification} reports classification accuracy on the DUKE Liver dataset.

\begin{table*}[ht]
\centering
\caption{The classification accuracy (ACC, \%) for sequence identification on DUKE Liver. Non-parametric bootstrapping with 1,000 bootstrap replicates is employed for statistical analysis, and we report the 95\% confidence intervals (CIs) in parentheses. For each dataset, the best-performing model is in bold, and the second-best-performing model is underlined. The $\Delta$ (Swin-B) metric quantifies the performance improvement of our pre-trained model over the Swin-B baseline trained from scratch. Significance marks (*, **, ***) indicate $p<0.05$, $p<0.01$, and $p<0.001$, respectively.}
\label{tab:sequence_identification}
\begin{tabular}{ll}
Model\textbackslash   Dataset & DUKE Liver                                   \\ \midrule
ResNet3D                      & 75.178 (68.837-81.519), p$<$0.001            \\
Swin-B                        & 74.406 (69.512-79.300), p$<$0.001            \\
MAE 3D                        & 76.760 (71.246-82.274), p$<$0.01             \\
MoCo V3                       & 77.011 (71.247-82.776), p$<$0.01             \\
SwinUNETR                     & \underline{77.835 (73.085-82.585)}, p$<$0.05 \\
BrainSegF.                    & 75.754 (70.129-81.379), p$<$0.01             \\
Ours                          & \textbf{78.210 (73.047-83.373)}              \\
$\Delta$ (Swin-B)             & 3.804 (***)                                    \\ \bottomrule
\end{tabular}
\end{table*}

\clearpage

\subsection{Report Generation}
We evaluated \modelname~on radiology report generation using a private knee dataset and assessed performance with three commonly used natural language generation metrics: BLEU (1-4), METEOR, and ROUGE-L. STable~\ref{tab:report} presents the quantitative results, where our model achieves the highest scores across nearly all metrics, including BLEU-1 to BLEU-4, METEOR, and ROUGE-L.
\begin{table*}[ht]
    \centering
    \caption{We report three major metrics for report generation, BLEU (1-4), METEOR, and ROUGE-L. Our model achieved an average improvement of 2.1\%. The $\Delta$ (Swin-B) metric quantifies the performance improvement of our pre-trained model over the Swin-B baseline trained from scratch. Significance marks (*, **, ***) indicate $p<0.05$, $p<0.01$, and $p<0.001$, respectively.}
    \label{tab:report}
    \resizebox{\textwidth}{!}{
    \begin{tabular}{lllllll}
        \toprule
    Model    & BLEU-1 & BLEU-2 & BLEU-3 & BLEU-4 & METEOR & ROUGE-L \\
        \midrule
    ResNet3D         & 30.65 (±1.56)          & 16.13 (±2.22)          & 6.54 (±3.19)          & 3.92 (±1.51)          & 8.62 (±1.91)           & 21.46 (±2.90)          \\
    Swin-B           & 27.97 (±1.09)          & 15.83 (±2.96)          & 6.68 (±1.33)          & 4.50 (±2.36)          & 9.05 (±2.84)           & 22.99 (±1.85)          \\
    MAE 3D           & 30.05 (±2.82)          & 13.70 (±2.77)          & 6.99 (±2.81)          & \textbf{5.41 (±1.92)} & 8.39 (±3.08)           & 22.50 (±2.87)          \\
    MoCo V3          & 28.69 (±3.43)          & 15.08 (±1.46)          & 6.74 (±1.80)          & 5.03 (±1.38)          & 9.16 (±2.59)           & 21.45 (±2.39)          \\
    SwinUNETR        & 30.40 (±2.71)          & 16.75 (±2.27)          & 7.92 (±1.29)          & 5.27 (±1.99)          & 10.42 (±1.15)          & 23.71 (±2.28)          \\
    BrainSegF.       & 29.54 (±2.07)          & 16.10 (±1.78)          & 7.33 (±2.89)          & 4.54 (±2.79)          & 10.64 (±1.42)          & 23.71 (±2.20)          \\
    Ours             & \textbf{30.84 (±1.45)} & \textbf{17.27 (±1.83)} & \textbf{9.29 (±2.35)} & 5.19 (±3.43)          & \textbf{11.19 (±2.64)} & \textbf{24.16 (±1.95)} \\
    \midrule
    $\Delta$(Swin-B) & 2.88 (**)               & 1.43 (*)               & 2.61 (**)             & 0.69 (**)             & 2.15 (***)              & 1.17 (**)            \\
    \bottomrule
    \end{tabular}
    }
\end{table*}

\clearpage
\subsection{Scalability and Ablation Study}
We conducted a comprehensive scalability and ablation study to assess the impact of pretraining dataset size and model capacity on downstream performance. As listed in STable~\ref{tab:scal_config}, this evaluation spans four representative tasks—sequence identification, age estimation, disease classification, and lesion/organ segmentation, using 10 datasets covering a range of anatomical regions and task types.
%STable~\ref{tab:scal_law} presents the performance comparisons across different pretraining dataset scales (10K, 53K, 336K) and model sizes (Swin-B, Swin-L, Swin-H). 

\begin{table*}[ht]
    \centering
    \caption{The datasets and corresponding tasks employed in the scalability evaluation and ablation study. The datasets and experimental configurations are maintained consistently with those used in other downstream evaluations.}
    \label{tab:scal_config}
    \begin{tabular}{lll}
    \toprule
    Dataset    & Task                    &  \\
    \midrule
    DUKE Liver & Sequence Identification &  \\
    ADNI       & Age Estimation          &  \\
    OASIS      & Age Estimation          &  \\
    OpenBHB    & Age Estimation          &  \\
    PI-CAI      & Disease Grading        &  \\
    MRNet      & Abnormality Diagnosis   &  \\
    AMOS       & Organ Segmentation      &  \\
    OAI-ZIB     & Organ Segmentation      &  \\
    PI-CAI      & Lesion Segmentation     &  \\
    BraTS 2021  & Lesion Segmentation    &  \\
    \bottomrule
    \end{tabular}
\end{table*}

\bibliographystyle{unsrt}
\bibliography{sn-bibliography}% common bib file